\documentclass[aps,prd,preprintnumbers,superscriptaddress,nofootinbib]{revtex4}
\usepackage{cleveref}
\usepackage{amsmath}
\usepackage{amssymb}
\usepackage{amsthm}
\usepackage{mathtools}
\usepackage{mathrsfs}
\usepackage{bm}
\usepackage{slashed} 
\usepackage{graphicx}
\usepackage{multirow}
\usepackage{tikz}
\usepackage[caption=false]{subfig}
\usepackage{relsize}	
\usepackage{array}
\usepackage{float}
\usepackage{color}
\usepackage{xcolor}
\usepackage{soul}
\usepackage{verbatim} 
\usepackage{siunitx}
\usepackage{hyperref}
\hypersetup{colorlinks=true,linkcolor=purple,anchorcolor=blue,citecolor=purple1, filecolor=blue,urlcolor=red,bookmarksnumbered=true,
pdfview=FitB
}
\allowdisplaybreaks

\def\be{\begin{eqnarray}}
\def\ee{\end{eqnarray}}
\colorlet{purple1}{blue!70!red}
\colorlet{darkred}{red!50!black}

\begin{document}
\title{Form Factors and Generalized Parton Distributions of Heavy Quarkonia \\ in Basis Light Front Quantization}
\author{Lekha~Adhikari}\email{adhikari@iastate.edu}
\affiliation{Department of Physics and Astronomy, Iowa State University,
Ames, IA 50011, U.S.A.}
\author{Yang~Li}\email{yli48@wm.edu}
\affiliation{Department of Physics, College of William \& Mary, Williamsburg, VA 23187, U.S.A.}
\author{Meijian~Li}\email{meijianl@iastate.edu}
\affiliation{Department of Physics and Astronomy, Iowa State University,
Ames, IA 50011, U.S.A.}
\author{James~P.~ Vary}\email{jvary@iastate.edu}
\affiliation{Department of Physics and Astronomy, Iowa State University,
Ames, IA 50011, U.S.A.}
\date{\today}
\begin{abstract}
We calculate the electromagnetic (charge, magnetic and quadrupole) form factors and the associated static moments of  heavy quarkonia (charmonia and bottomonia) using the Basis Light Front Quantization (BLFQ) approach. For this work, we adopt light front wavefunctions (LFWFs)  generated by a holographic  QCD confining potential and a one-gluon exchange interaction with fixed coupling. We compare our BLFQ results with the limiting case of a single BLFQ basis state description of heavy quarkonia and with other available results. These comparisons provide insights into relativistic effects. Using the same LFWFs generated in the BLFQ approach, we also present the generalized parton distributions (GPDs) for selected mesons including those for radially excited mesons such as $\psi^\prime$ and $\Upsilon^\prime$. Our GPD results establish the foundation within BLFQ for further investigating hadronic structure such as probing the spin structure of spin-one hadrons in the off-forward limit.
\end{abstract}
\maketitle
\section{Introduction}
Exploring the electromagnetic (EM) properties of  spin-one hadrons has been of great interest because it provides insight into the spin-sensitive structure and the internal dynamics of the hadrons. In particular, hadronic form factors (FFs) serve as one important tool to understand the structure of bound states in quantum chromodynamics (QCD). The numerous investigations on the structure of  the spin-zero and spin-one hadrons that include FFs in different formalisms ~\cite{deMelo:1997hh, Roberts:2011wy, Dudek:2006ej, Maris:2006ea, Karmanov:1996qc,Cardarelli:1994yq, Bakker:2002mt,Hilger:2017jti,Bedolla:2018lif,Bhagwat:2006pu,Chung:1988my,Frankfurt:1993ut, Grach:1983hd,Raya:2017ggu,Bedolla:2016yxq,Bedolla:2015mpa,Brodsky:1992px,Brodsky:2007hb,Hawes:1998bz,Li:2017uug,Li:2016wwu,Li:2017mlw,Li:2015zda,Swarnkar:2015osa} provide a window for understanding hadronic structure at low and medium momentum transfer. The investigations  with relativistic approaches~\cite{deMelo:1997hh, Karmanov:1996qc,Swarnkar:2015osa, Bakker:2002mt,Brodsky:1992px, Brodsky:2007hb,Grach:1983hd,Cardarelli:1994yq,Li:2017uug,Li:2016wwu,Li:2017mlw,Li:2015zda,Chung:1988my,Frankfurt:1993ut} have presented results for FFs, decay constants and the distribution amplitudes of spin-zero and spin-one bound-state systems such as the pion ($\pi$), kaon ($K$), rho meson ($\rho$) and ~$J/\psi$ meson adopting different light front (LF) models. Note a recent investigation~\cite{Li:2017uug} has shown the FFs of (pseudo) scalar mesons calculated in a general frame. That work has also pointed out the differences among the results calculated in the various frames including the Drell-Yan frame.

Despite these numerous studies and growing interests, there is little consensus on how to obtain static moments such as the quadrupole moments on the LF. Furthermore, investigations on the EM FFs and the associated static moments of radially excited vector mesons such as $\psi^\prime$ and $\Upsilon^\prime$ are rare to the best of our knowledge~\cite{Bedolla:2018lif,Hilger:2017jti}. It is therefore natural to ask what one can learn about the spin-one hadronic structure from an investigation using the recently developed non-perturbative Basis Light Front Quantization (BLFQ) approach. In this paper, we present the EM FFs and the associated static moments, the charge radii, magnetic moments and the quadrupole moments, for a selection of heavy quarkonia. We also calculate corresponding quantities in a simplified basis, which is referred to as the Single Basis Limit (SBL) approach in this paper, and interpret the differences with the BLFQ results to uncover dynamical effects arising from the different interactions. We also compare our results with corresponding results from the Contact Interaction (CI)~\cite{Bedolla:2015mpa,Raya:2017ggu,Bedolla:2016yxq}, Dyson-Schwinger Equation (DSE)~\cite{Maris:2006ea,Bhagwat:2006pu} and Lattice~\cite{Dudek:2006ej} approaches wherever available.

The generalized parton distribution (GPD) has emerged as a powerful tool to describe hadrons in terms of quark and gluon degrees of freedom. Several reviews cover the GPDs and their connections to experiments~\cite{Brodsky:2006in,Taneja:2011sy,Diehl:2003ny,Ji:PRL,mb:GPD,Berger:2001zb,Sun:2017gtz, Cano:2003ju, Frederico:2009fk, Liuti:2014dda,Mondal:2017lph, Choi:2001fc,Airapetian:2001yk}. In particular, there are several investigations on the spin-zero and spin-one GPDs \cite{Taneja:2011sy,Berger:2001zb,Sun:2017gtz, Cano:2003ju, Frederico:2009fk, Liuti:2014dda,Mondal:2017lph, Choi:2001fc}. For example, in Refs.~\cite{Frederico:2009fk, Choi:2001fc}, the pion GPDs have been calculated in the LF phenomenological models with both the valence and non-valence contributions. Similarly, in Ref.~\cite{Taneja:2011sy}, the angular momentum sum rule for the spin-one system within the gauge-invariant decomposition framework  has been investigated. That study has also discussed the connections between the deeply virtual Compton scattering (DVCS) amplitudes and the total quark angular momentum for the ground-state vector meson GPDs. In Ref.~\cite{Cano:2003ju}, The deuteron GPDs have been investigated within the impulse approximation in a framework with non-zero longitudinal momentum transfer. The recent study of the deuteron  in Ref.~\cite{Mondal:2017lph} has also presented the GPDs in the framework of holographic QCD.

These studies, however, do not address the heavy mesons (for example, $c{\bar c}$ and $b {\bar b}$). Hence, such basic properties as the momentum-transfer dependence of the valence quarks for the radially excited states $\psi^\prime$ and $\Upsilon^\prime$ are not previously available. Furthermore, we are motivated by the feasibility of experiments to investigate the hadronic structure of the (pseudo) scalar and vector meson GPDs in the forward limit (zero momentum transfer limit). Such measurements provide connections with unpolarized parton distributions~\cite{Berger:2001zb}.

In this work, we calculate the EM FFs and  GPDs through the corresponding matrix elements which are defined by the overlap integrals of light front wavefunctions (LFWFs) in the Drell-Yan frame. The non-perturbative solutions for the LFWFs are provided by a recent BLFQ study \cite{Li:2015zda} of heavy quarkonia. This work implements a transverse confining potential from light front holography and a longitudinal confining interaction which has a similar shape in the non-relativistic limit. It also includes the one-gluon exchange interaction with  a fixed coupling to generate the spin structure of the charmonium and bottomonium systems. Note a recent investigation ~\cite{Leitao:2017esb} following the quarkonia study~\cite{Li:2015zda} with a running coupling~\cite{Li:2017mlw} provides comparisons of the mass spectrum and decay constant between the results obtained from the BLFQ  and that from the covariant spectator theory (CST). The CST treatment~\cite{Leitao:2017esb} is an independent, fully relativistic approach and the comparison of the results of CST and BLFQ showed favorable correspondence.

For this work, we adopt the BLFQ approach which is developed for solving bound-state problems in quantum field theory \cite{Vary:2009gt, Wiecki.2014,Li:2015zda}. This approach not only provides easy conversion between the transverse coordinates and momentum space~\cite{Li:2017mlw,Adhikari:2016idg}, but also connects mass spectroscopy with other observables~\cite{Chen:2016dlk,Li:2015zda}. The BLFQ is a Hamiltonian-based formalism  that uses the advantages of LF dynamics \cite{Brodsky:1997de} with advances in solving many body bound-state problems \cite{Navratil:2000ww}. It has been successfully applied to the single electron problem in quantum electrodynamics (QED)~\cite{Zhao:2014xaa}, the strong coupling bound-state positronium problem ~\cite{Wiecki.2014,Adhikari:2016idg} and the running coupling quarkonium problem~\cite{Li:2017mlw} . Furthermore, the BLFQ approach has been extended to time-dependent strong external field problems such as non-linear Compton scattering~\cite{Zhao:2013jia}. The reviews related to BLFQ and its application are available in Refs.~\cite{Leitao:2017esb,Adhikari:2016idg,Li:2017mlw,Navratil:2000ww,Zhao:2014xaa,Chen:2016dlk,Wiecki.2014,Li:2018uif,
Vary:2016emi,Vary:2016ccz,Li:2015zda,Li:2017uug,Li:2016wwu,Vary:2009gt,Zhao:2013jia,Vary:2018pmv}.

We organize this paper as follows: In Sec.~\ref{ff_gpd}, FFs and GPDs are defined through the local and non-local matrix elements of the plus component of the current operator, respectively. Then, the FFs and GPDs are expressed in terms of the overlap integrals of LFWFs. In Sec.~\ref{formalism}, we briefly introduce BLFQ along with SBL, the simplified case of a single BLFQ basis state, and in Sec.~\ref{Results}, we present our results. Finally, we present the summary of this work and outlook for further research in Sec.~\ref{Summary}.
\section{Form factors and generalized parton distributions for spin-one hadrons on the light front \label{ff_gpd}}
The Lorentz-invariant, elastic FFs $F_i(t)$ for  spin-one hadrons are defined by the local matrix elements of the current operator $J^\mu [\triangleq \bar{\psi}(0)\gamma^\mu  \psi(0)$] as \cite{deMelo:1997hh,Lorce:2009bs,Bakker:2002mt}
\begin{equation}
\begin{split}
I_{m_J,m_J'}(t) &\triangleq \frac{1}{(p+p^\prime)^\mu} \langle p^\prime,J=1, m_J^\prime| J^\mu|p,J=1, m_J \rangle  \\
&= \frac{1}{P^\mu}\bigg[ - F_1(t)(\epsilon^{\prime *} \cdot \epsilon) P^\mu + F_2(t)[\epsilon^\mu(\epsilon^{\prime*}\cdot P)+\epsilon^{\prime * \mu} (\epsilon \cdot P)] - F_3(t)\frac{(\epsilon \cdot P)(\epsilon^{\prime * }\cdot P)}{2 M^2}P^\mu\bigg ],
\end{split}
\label{eq:ff_decomp}
\end{equation}
where $\psi (\bar{\psi})$ is the quark (anti-quark) field operator,  $p$ ($p^\prime$) is the momentum of the initial (final) state of the hadron, $J$ is total angular momentum for the hadron, $m_J$ ($m_J^{\prime}$) is the total angular-momentum projection in the initial (final) state of the hadron, $t\equiv(p^\prime-p)^2$, $M$ is the mass of the hadron,  $P= p^\prime +p$, $\epsilon = \epsilon(p,m_J)$ and $\epsilon^\prime= \epsilon^{\prime} (p^\prime , m_J^\prime)$ are the polarization vectors of the hadron in the initial and final helicity states, respectively, satisfying $\epsilon \cdot p= \epsilon^\prime \cdot p^\prime =0$, and $I_{m_J,m_J'}(t)$ represents the helicity amplitudes. In this work, the possible values $+1, -1$ and $0$ of $m_J$ (and $m_J'$) for the spin-one hadrons are represented by $+,\,-$ and $0$, respectively. For simplicity, the charge of the quark is excluded in the definition of the FFs in Eq.~\eqref{eq:ff_decomp}.

We adopt the following conventions~\cite{Bakker:2002mt} to calculate helicity amplitudes $I_{m_J,m_J'}(t)$ in the Drell-Yan equivalent frame.
\begin{equation}
\begin{split}
p^\mu = (\frac{M\sqrt{1+\tau}}{\sqrt{2}}, \frac{M\sqrt{1+\tau}}{\sqrt{2}}, -\frac{q}{2}, 0), \quad p^{\prime \mu} = (\frac{M\sqrt{1+\tau}}{\sqrt{2}}, \frac{M\sqrt{1+\tau}}{\sqrt{2}}, \frac{q}{2}, 0),\\
\epsilon^\mu(p,m_J =\pm)= \mp \frac{1}{\sqrt{2}}(0, -\frac{q}{2p^+}, 1,\pm i), \quad \epsilon^\mu(p^\prime,m_J^\prime =\pm)= \mp \frac{1}{\sqrt{2}}(0, \frac{q}{2p^+}, 1,\pm i), \\
\epsilon^\mu(p,0) = \frac{1}{M}(p^+, \frac{-M^2 + q^2/4}{2p^+}, -\frac{q}{2}, 0), \quad {\epsilon'}^\mu(p,0) = \frac{1}{M}(p^+, \frac{-M^2 + q^2/4}{2p^+}, \frac{q}{2}, 0),
\end{split}\label{eq:spinor_ff}
\end{equation}
where $v^\mu\equiv (v^+, v^-, v^x, v^y)$ is the light front variables in this paper, $q=\sqrt{-t}$ and $\tau \equiv-q^2/(4 M^2)$.

There is only one helicity amplitude $I_{0, 0}(t)$ for $J=m_J=0$ that can be computed from the plus component of the current defined in Eq.~\eqref{eq:ff_decomp}, and the charge form factor for spin-zero hadron is therefore defined by $G_C(t)\equiv I_{0, 0} (t)$. But, in the case of  $J=1$ with $m_J$ (and $m_J'$ )$=+, 0, -$, there are nine helicity amplitudes $I_{m_J,m_J'}(t)$ that can be computed for the same current. One can reduce them to four amplitudes $I_{+, -}(t)$, $I_{+, +}(t)$, $I_{+, 0}(t)$ and $I_{0, 0}(t)$ using the light front parity and the charge conjugation symmetries in LF dynamics.

Using Eq.~\eqref{eq:spinor_ff} in Eq.~\eqref{eq:ff_decomp}, it is straightforward to extract the four  helicity amplitudes as
\begin{equation}
\begin{split}
 &I_{+,+}(t) = F_1(t)+\tau F_3(t), \quad{I_{+,0}(t) = \sqrt{\frac{\tau}{2}} \big[2F_1(t) -F_2(t) + 2 \tau F_3(t)\big],}\\
&I_{+,-}(t)= - \tau F_3(t), \quad{I_{0,0}(t) =(1-2 \tau)F_1(t) +2 \tau F_2(t) - 2\tau^2 F_3(t)}.
\end{split}
\end{equation}
In the case of the spin-one hadrons, there are  three Lorentz-invariant, elastic FFs $F_i(t)$, and hence three EM FFs, but there are four helicity amplitudes. Studies presented in Refs.~\cite{deMelo:1997hh, Brodsky:1992px, Chung:1988my, Grach:1983hd, Bakker:2002mt, Karmanov:1996qc} have claimed that computing the EM FFs is more feasible than the elastic FFs $F_i(t)$. The four helicity amplitudes in LF dynamics and the three EM FFs create an ambiguity on how to compute them in the case where the current conservation is not preserved, and therefore the relations that define the EM FFs from the helicity amplitudes are not unique. There are several choices in which the four helicity amplitudes can be combined  to extract the EM FFs. One can find the most popular choices in Refs.~\cite{deMelo:1997hh, Chung:1988my, Grach:1983hd, Brodsky:1992px,Frankfurt:1993ut}. The studies available in Refs.~\cite{Bakker:2002mt, Karmanov:1996qc, deMelo:1997hh} suggest to adopt the prescription defined by Grach and Kondratyuk (GK) available in Refs.~\cite{Grach:1983hd, Cardarelli:1994yq} because this prescription does not contain any contribution from the helicity amplitude $I_{0,0}$(t) showing the prescription free from the zero-mode contributions. In this work, we therefore use the GK prescription to calculate EM FFs.

Following the GK prescription, one can define the three EM FFs, the charge FF $G_C(t)$, the magnetic FF $G_M(t)$ and the quadrupole FF $G_Q(t)$, in terms of the four helicity amplitudes as
\be
 G_C(t)&=& \frac{1}{3}  \big[ (3- 2 \tau)I_{+, +}(t) + I_{+, -}(t) +2 \sqrt{2 \tau} I_{+, 0}(t)\big], \label{eq:c_amplitude} \\
G_M(t)&=&  2I_{+, +}(t) -\sqrt{\frac{2}{\tau}} I_{+, 0}(t),   \label{eq:m_amplitude} \\
 G_Q(t)&=& \frac{2\sqrt{2}}{3}  \big[ -\tau I_{+, +}(t) - I_{+, -}(t) +\sqrt{2\tau} I_{+, 0}(t) \big ]. \label{eq:q_amplitude}
\ee
The charge root-mean-squared (r.m.s.) radius $\sqrt{\langle r^2 \rangle}$, magnetic moment $\mu$  and the quadrupole moment {\text Q} are defined by~\cite{Chung:1988my}
\be
\langle r^2 \rangle&=&-6\frac{\partial}{\partial t}G_C(t)\biggr |_{t\rightarrow 0}, \label{eq:charge_radius} \\
\mu &=&G_M(t=0), \label{eq:magnetic_moment}\\
{\text Q} &=& 3\sqrt{2}\frac{\partial}{\partial t}G_Q(t) \biggr |_{t\rightarrow 0} \label{eq:quadrupole_moment}
\ee
 with normalization $G_C(t=0) =1$. Note that heavy quarkonium is charge symmetric, thus the total charge of the system is zero. We therefore calculate the form factors by considering only ``the quark" contribution. Although this case is fictitious, it is well-defined and  can be compared with related spin-one-hadron theoretical work. It is noted that, in LFWF representation~\cite{Brodsky:2000ii}, the r.m.s. radii can also be related to the impact parameter ${\mathbf b}_\perp \equiv (1-x) {\mathbf r}_\perp$~\cite{mb:GPD}  by $\langle r^2 \rangle = (3/2) \langle { b}_\perp^2 \rangle$~\cite{Li:2017mlw, Li:2016wwu}.

One can define a total of nine real GPDs for the spin-one hadrons through the non-local matrix elements of the (axial) vector current on the LF. Five of them are computed from the non-local matrix elements of the same current operator (plus component) which is used as a local operator in Eq.~\eqref{eq:ff_decomp} whereas the remaining four are computed from that of the axial current \cite{Berger:2001zb,Cano:2003ju}. Although there are nine non-local matrix elements that can be computed from the plus component of the current, only five of them are linearly independent because of the constraints from parity invariance. Thus, there are  five real GPDs that can be calculated from the five linearly independent non-local matrix elements. In this paper, we only present  the (pseudo) scalar and vector meson GPDs that are computed from the current with no quark helicity flip because the meson GPDs with no helicity flip are the ones most readily compared with  phenomenological applications~\cite{Berger:2001zb}. It is however straightforward to calculate helicity-flip GPDs using the same method that is used for helicity-non-flip GPDs.

The five vector meson GPDs for the spin-one hadron  are defined through the non-local matrix elements of the vector current on the LF as \cite{Berger:2001zb,Cano:2003ju}
\begin{equation}
\begin{split}
V_{m_J, m_J^\prime}(x,\xi, t)& \triangleq \int \, \frac{dz^-}{2 \pi} e^{ix P^+ z^-}\langle p^\prime, J=1, m_J^\prime| \bar{\psi}\big(-\frac{z^-}{2}\big)\gamma^+  \psi\big(\frac{z^-}{2}\big)|p,J=1, m_J \rangle \biggr |_{z^+=0,\mathbf{z_\perp} = \mathbf{0}_\perp} \\
& =-(\epsilon^{\prime *}\cdot \epsilon) H_1(x, \xi, t) + \frac{(\epsilon \cdot n)(\epsilon^\prime \* \cdot P)+(\epsilon^{\prime *}\cdot n)(\epsilon \cdot P)}{P\cdot n} H_2(x, \xi, t) -\frac{(\epsilon \cdot P)(\epsilon^{\prime * }\cdot P)}{2 M^2} H_3(x, \xi, t) \\
&+\frac{(\epsilon \cdot n)(\epsilon^\prime \* \cdot P)-(\epsilon^{\prime *}\cdot n)(\epsilon \cdot P)}{P\cdot n} H_4(x, \xi, t) + \biggr [ 4 M^2 \frac{(\epsilon \cdot n)(\epsilon^{\prime *} \cdot n)}{(P\cdot n)^2}  \biggr ] H_5(x, \xi, t).
\end{split}
\label{eq:gpd_decomp}\end{equation}
Here, $n = (1, 0, 0, 1)$ is a null vector perpendicular to the light front direction. We choose Ji's convention \cite{Ji:1998pc} to define arguments $x$, $\xi$ and $t$ of the GPDs $H_i$, $i=1,2,\dots,5$, where $x$ is the momentum fraction carried by the quark in the longitudinal direction and $\xi$ is the skewness parameter. In this work, we choose the Drell-Yan frame $\Delta^+=0$, or equivalently $\xi =0$ so that $\Delta^2 (\equiv t ) = - \mathbf{\Delta}_\perp^2 <0$. 

It is straightforward to extract the five GPDs in terms of the five linearly independent non-local matrix elements using Eq.~\eqref{eq:spinor_ff} in Eq.~\eqref{eq:gpd_decomp}. Due to the time reversal symmetry on matrix elements $V_{m_J, m_J^\prime}(x, 0, t)$, one can write $V_{+,0}(x, 0, t)= - V_{0,+}(x, 0, t)$ \cite{Cano:2003ju,Berger:2001zb}, and we therefore choose those independent non-local matrix elements to be $V_{0,0}(x, 0, t), \,V_{+,+}(x, 0, t),\,V_{+,0}(x, 0, t), \,V_{+,-}(x, 0, t)$. Thus, the expressions for the GPDs read
\be
H_1(x, 0, t) &=& \frac{1}{3} [V_{0,0}(x, 0, t)- 2(\tau -1)V_{+,+}(x, 0, t)+ 2 \sqrt{2\tau} V_{+,0}(x, 0, t)+2 V_{+,-}(x, 0, t)], \label{eq:gpd_amplitude1}\\
H_2(x, 0, t) &=& 2 V_{+,+}(x, 0, t) - \frac{2}{\sqrt{2\tau}}V_{+,0}(x, 0, t),\label{eq:gpd_amplitude2} \\
H_3(x, 0, t) &=& -\frac{V_{+,-}(x, 0, t)}{\tau}, \label{eq:gpd_amplitude3}\\
H_4(x, 0, t) &=& 0, \label{eq:gpd_amplitude4}\\
H_5(x, 0, t) &=& V_{0,0}(x, 0, t)- (1+2 \tau) V_{+,+}(x, 0, t) + 2\sqrt{2 \tau} V_{+,0}(x, 0, t)-V_{+,-}(x, 0, t).
\label{eq:gpd_amplitude5}\ee

Note our expressions are consistent with those from Ref.~\cite{Cano:2003ju} in the limit $\xi=0$.  It is interesting to observe that the integrations of $H_4(x, 0, t)$ and $H_5(x, 0, t)$ over $x$ do not correspond to $F_i(t)$  of the local current [see Eq.~\eqref{eq:ff_decomp}] and therefore vanish. This arises from the time reversal constraints in the case of $H_4(x, 0, t)$. In the case of $H_5(x, 0, t)$, this arises because of the term $n^\mu n^\nu/(P \cdot n)^2$ whose analog is absent in the decomposition of the local current [Eq.~\eqref{eq:ff_decomp}] as a consequence of Lorentz-invariance \cite{Berger:2001zb,Cano:2003ju}. Here, we point out that the right-hand side of Eq.~\eqref{eq:gpd_amplitude5}, after integrating over $x$, is widely known and cited in the spin-one FF calculations as  the angular condition~\cite{Grach:1983hd,Bakker:2002mt}.  Thus, the first moments of  the GPDs can be related to $F_i(t)$ for the spin-one hadrons  by the first set of sum rules on the LF as \cite{Berger:2001zb,Cano:2003ju}
\be \int   H_i(x, 0, t)\, dx = F_i(t), \quad{i=1, 2, 3}, \label{eq:H123_ff}\ee
\be
\int  H_4(x,0, t)\, dx = 0,\quad{\int H_5(x, 0, t)\, dx = 0}.\label{eq:H45zero}
\ee
Similarly, the second moments of the GPDs can be related to gravitational FFs by a second set of sum rules (via stress tensor decomposition) as defined in Refs.\cite{Abidin:2008ku,Taneja:2011sy}.

In the Drell-Yan frame, within the impulse approximation, the helicity amplitudes $I_{m_J, m_J^\prime}(t)$ and the non-local matrix elements $V_{m_J, m_J^\prime}(x, 0, t)$  in the region $0\leq \!x\! \leq \!1$ can be written as overlap integrals between LFWFs. The expression for $V_{m_J, m_J^\prime}(x, 0, t)$ reads ~\cite{Diehl:2003ny,Frederico:2009fk}
\begin{equation}
\begin{split}
 V_{m_J,m_J'}(x,0, t) =  \sum_{\lambda_q,\lambda_{\bar q}}
 \, \int \frac{d^2 {\mathbf k}_\perp}{2x(1-x)(2\pi)^3} \,  \psi_{m'_J}^{J*}(\mathbf k'_\perp, x, \lambda_q,\lambda_{\bar q})\,  \psi_{m_J}^{J}(\mathbf k_\perp, x, \lambda_q,\lambda_{\bar q}) \label{eq:GPDs_LFWF}
\end{split}
\end{equation}
and that for $I_{m_J, m_J^\prime}(t)$ reads \cite{Brodsky:2007hb,Li:2017uug,Drell:1969km}
\begin{equation}
\begin{split}
 I_{m_J,m_J'}(t) =  \sum_{\lambda_q,\lambda_{\bar q}}
\int_0^1 \frac{dx}{2x(1-x)}\, \int \frac{d^2 {\mathbf k}_\perp}{(2\pi)^3} \,  \psi_{m'_J}^{J*}(\mathbf k'_\perp, x, \lambda_q,\lambda_{\bar q})\,  \psi_{m_J}^{J}(\mathbf k_\perp, x, \lambda_q,\lambda_{\bar q}), \label{eq:ff_LFWF}
\end{split}
\end{equation}
where ${\mathbf k}_\perp$ and  ${\mathbf k}^\prime_\perp = {\mathbf k}_\perp +(1-x) \mathbf {\Delta}_\perp $ are the respective relative transverse momenta of the quark before and after being struck by the virtual photon, $\lambda_q(\lambda_{\bar{q}})$  is the helicity of the quark (anti-quark). Note that integrating $V_{m_J, m_J^\prime}(x, 0, t)$ over $x$ yields the local matrix elements (helicity amplitudes) $I_{m_J, m_J^\prime}(t)$. Here, the LFWFs are truncated to only the valence Fock sector. The valence sector LFWF  is normalized according to \cite{Li:2015zda}
\begin{equation}
\begin{split}
\sum_{\lambda_q,\lambda_{\bar q}}
\int_0^1 \frac{dx}{2x(1-x)}\, \int \frac{d^2 \mathbf k_\perp}{(2\pi)^3} \, \biggr | \psi_{m_J}^{J}(\mathbf k_\perp, x, \lambda_q,\lambda_{\bar q})\biggr |^2 =1. \label{eq:normalization_LFWF}
\end{split}
\end{equation}
Note that this convention for normalization is introduced in Ref.~\cite{Li:2015zda} and the non-perturbative solutions of LFWFs are generated accordingly.
\section{formalism \label{formalism}}
\subsection{Basis Light Front Quantization (BLFQ) \label{BLFQ}}
A recent study of heavy quarkonia \cite{Li:2015zda}, in a LF Hamiltonian approach \cite{Vary:2009gt}, presents the effective Hamiltonian based, in part, on  the LF holographic QCD \cite{Brodsky:2014yha} as
\be H_{\text{eff}} \equiv \frac{{\mathbf k}_\perp^2 +{\text m}^2_q}{x(1-x)} +V_T +V_L + V_g,\label{eq:H_eff}
\ee
where ${\text m}_q$ is the mass of the quark. $V_T$ is the ``soft-wall" light front holography  in the transverse direction and is defined as
\begin{equation}
V_T \equiv \kappa^4 {\bold \zeta}^2_\perp = \kappa^4 x(1-x) {\mathbf r}_\perp \quad{}\text{with}  \quad{{\mathbf r}_\perp = \mathbf {r}_{q \perp}-{\mathbf r}_{\bar q \perp}},
\end{equation}
where $\bold \zeta$ is holographic variable \cite{Brodsky:2014yha}, $\kappa$ is the confining strength, and ${\mathbf r}_\perp$ is the transverse separation between the quark and the anti-quark. The longitudinal confining potential reads
\begin{equation} V_L \equiv -\frac{\kappa^4}{(2{\text m_q})^2}\partial_x(x(1-x)\partial_x)\quad{} \text{with} \quad{\partial_x \equiv (\partial/\partial x)_{\zeta_{\perp}}}.
\end{equation}
$V_g$ is the one-gluon exchange term and in the momentum space, it reads \cite{Wiecki.2014,Li:2015zda}
\begin{equation}
V_g = - \frac{C_F 4\pi \alpha_s}{Q^2}\bar{u}_{\lambda^\prime_q}({\text k}^\prime)\gamma_\mu u_{\lambda_q}({\text k})\bar{v}_{\lambda_{\bar{q}}}(\bar{{\text k}})\gamma^\mu v_{\lambda^\prime_{\bar{q}}}(\bar{{\text k}}^\prime),
\end{equation}
where $C_F=4/3$ is the color factor for the color singlet state, $\alpha_s$ is the fixed coupling constant, and $Q^2= -(1/2) ({\text k}^\prime - {\text k})^2 -(1/2)(\bar{{\text k}}^\prime - \bar{{\text k}})^2$ is the average momentum squared carried by the exchanged gluon.

In the BLFQ approach, if  quarkonium is described  by state vectors $|\psi^J_{m_J}\rangle$, the eigenvalue equations can be defined by
\be H_{\text{eff}}|\psi^J_{m_J}\rangle= M^2|\psi^J_{m_J}\rangle \label{eq:Heff_eigen_value} \ee
and solved non-perturbatively to obtain eigenfunctions  that represent the LFWFs $\psi^J_{m_J}({\mathbf  k}_\perp, x, \lambda_q, \lambda_{\bar q})$ for heavy quarkonium.
To solve Eq.~\eqref{eq:Heff_eigen_value}, two functions $\phi_{nm}$ and $\chi_l$ are adopted to form the basis in which to evaluate the Hamiltonian matrix. In the transverse direction, 2-dimensional (2D) harmonic oscillator (HO) functions are adopted and are defined, in terms of the dimensionless transverse momentum variable ${\bf v}_\perp$ ($= {\mathbf k}_\perp/b$), by~\cite{Li:2015zda}
 \begin{equation}
 \phi_{nm}({\bf v}_\perp) = e^{i m \theta} v^{|m|} e^{- v^2/2} L_n^{|m|}(v^2),\label{eq:transverse_basis}
\end{equation}
 where $v = |{\mathbf v}_\perp|$, $\theta = \arg {\mathbf v}_\perp$,
 $n$ and $m$ are the radial and angular quantum numbers,
 $L_n^{|m|}(z)$ is the associated Laguerre polynomial and $b$ is the HO basis scale with dimension of mass.
 In the longitudinal direction, the basis functions are defined by
  \begin{equation}
 \chi_l(x;\alpha,\beta) = \sqrt{4\pi(2l+\alpha +\beta +1)}\sqrt{\frac{\Gamma(l+1) \Gamma(l+\alpha +\beta +1)}{\Gamma(l+\alpha +1) \Gamma(l+\beta +1)}} x^{\frac{\beta}{2}} (1-x)^{\frac{\alpha}{2}}P_l^{(\alpha, \beta)}(2x-1),\label{eq:longitudinal_basis}
 \end{equation}
 where $P_l^{(\alpha, \beta)}(z)$ is the Jacobi polynomial, $\alpha=\beta =4{\text m}_q^2/\kappa^2$ are dimensionless basis parameters, and we drop $\alpha$ and $\beta$ from the arguments of $\chi_l$ hereafter.

 Using Eqs.~\eqref{eq:transverse_basis} and~\eqref{eq:longitudinal_basis} as basis functions, the expansion of momentum-space LFWFs reads \cite{Adhikari:2016idg,Li:2015zda}
  \begin{equation}
\psi^J_{m_J}({\mathbf k}_\perp, x, \lambda_q, \lambda_{\bar q}) =
\frac{1}{b} \sum_{n, m, l} \langle n, m, l, \lambda_q, \lambda_{\bar q} | \psi^J_{m_J}\rangle
\phi_{nm} \left( \frac{{\mathbf k}_\perp}{b \sqrt{x(1-x)}} \right) \chi_l(x), \label{eq:LFWF_blfq}
\end{equation}
where $\langle n, m, l,\lambda_q,\lambda_{\bar q}|\psi^J_{m_J}\rangle$ are the LFWFs in the BLFQ basis, obtained by diagonalizing the truncated Hamiltonian matrix~\cite{Li:2015zda}. The following truncation is applied to restrict the quantum numbers.
  \be 2n +|m|+1 \leq N_{\text{max}}, \quad{ l\leq L_{\text{max}} .} \label{eq:truncation1}\ee
 It is clear from the truncation  that $L_{\text{max}}$ controls the basis resolution in the longitudinal direction whereas $N_{\text{max}}$ controls the transverse momentum covered by 2D-HO functions. In the BLFQ approach, the total angular momentum $J$ is only an approximate quantum number due to the breaking of the rotational symmetry by the Fock sector truncation and the basis truncation. However, the total angular momentum projection $(m_J)$ for the system is conserved.
\be  m_J = m + \lambda_q  + \lambda_{\bar{q}}\label{eq:truncation2}.\ee

Inserting Eq.~\eqref{eq:LFWF_blfq} in Eqs.~\eqref{eq:ff_LFWF} and~\eqref{eq:GPDs_LFWF} yields the integral over the product of the two 2D-HO functions with different arguments, and that is simplified using the TM coefficients \cite{Talmi.1952} to reduce it to an integral over one 2D-HO function~\cite{Adhikari:2016idg}. Then the integral is calculated numerically. Readers are referred to Refs.~\cite{Li:2015zda, Adhikari:2016idg} for further details of the BLFQ approach.
\subsection{Single Basis Limit (SBL) \label{SBL}}
We investigate a special limiting case of BLFQ for calculating EM FFs of heavy quarkonia. For this purpose, we select the leading basis function contribution(s) of the LFWFs and scale them to become the sole normalized LFWF for the state in question. We refer to this severely limited basis space as the Single Basis Limit (SBL). The SBL represents an eigenstate of the Hamiltonian with the omission of the one-gluon exchange term. Thus, the difference between results with the BLFQ for the LFWFs and the SBL results provides insights into the role of configuration mixing induced by the effective one-gluon exchange interaction. Where the differences in a given observable are large we surmise that the gluon-exchange dynamics plays a significant role.

\section{Results and discussion \label{Results}}
In this section, we present and discuss our results for FFs, associated static moments and GPDs. The details of the Hamiltonian's parameters used in calculations are summarized in Table~\ref{tab:model_parameters}. The fixed gluon mass $\mu_g = 0.02\, {\text{GeV}}$ is introduced  to regularize the singularity present in Eq.~\eqref{eq:Heff_eigen_value}~\cite{Li:2015zda}. The convergence study of mass eigenvalues with different $\mu_g$ keeping $N_{\text{max}}= L_{\text{max}}$ fixed in Ref.~\cite{ Li:2015zda} suggested that the mass eigenvalues are well converged with respect to $\mu_g$. Therefore, the gluon mass is kept fixed in these calculations. Similarly, the HO basis scale $b$ is chosen to be equal to the confining strength $\kappa$ at the given $N_{\text{max}}= L_{\text{max}}$ value and at the fixed gluon mass $\mu_g$. Fixed, but flavor-dependent, coupling constants $\alpha_s$ are used to produce results presented in this work.

Our masses are obtained from the mass eigenvalue equations for total angular momentum projection  $m_J = 0$ at the given  $N_{\text{max}}= L_{\text{max}}$ truncation.  An important issue that arises in a LF Hamiltonian approach, such as BLFQ, concerns the relative sign  between different eigenstates. In particular, since the relative sign between two states with different $m_J$ is not fixed by the diagonalization (though the signs of all basis states are fixed by our basis state conventions), we control the overall sign of each eigenfunction to have positive derivative at the origin in coordinate space.
\begin{table}
\caption{Summary of the model parameters~\cite{Li:2015zda}.
}\label{tab:model_parameters}
 \centering
\begin{tabular}{ccc ccc ccc c}

\toprule
 meson\, &\, $N_{\text{max}} (=L_{\text{max}})$ \, & \, $\alpha_s$ \,& $ \mu_g (\text {GeV})$ \, & \, $\kappa (\text {GeV})$  \,& $m_q (\text {GeV})$\,  &    \\

\colrule

\multirow{1}{*}{} & \multirow{1}{*}{8} &0.3595  &0.02  & 0.963 & 1.49  &  \\
\multirow{1}{*}{$c\bar{c}$} & \multirow{1}{*}{16} &0.3595  &  0.02& 0.950 & 1.51 &    \\
\multirow{1}{*}{} & \multirow{1}{*}{24} & 0.3595 & 0.02 & 0.938 & 1.52 &   \\
\colrule
\colrule

\multirow{1}{*}{} & \multirow{1}{*}{8} &0.2500  &0.02  & 1.422 & 4.77 &   \\
\multirow{1}{*}{$b\bar{b}$} & \multirow{1}{*}{16} &0.2500  &  0.02& 1.423 & 4.78 &   \\
\multirow{1}{*}{} & \multirow{1}{*}{24} & 0.2500 & 0.02 & 1.422 & 4.78 &   \\

\botrule
\end{tabular}
\end{table}

\begin{figure*}
\begin{tabular}{cc}
\subfloat{\includegraphics[scale=0.48]{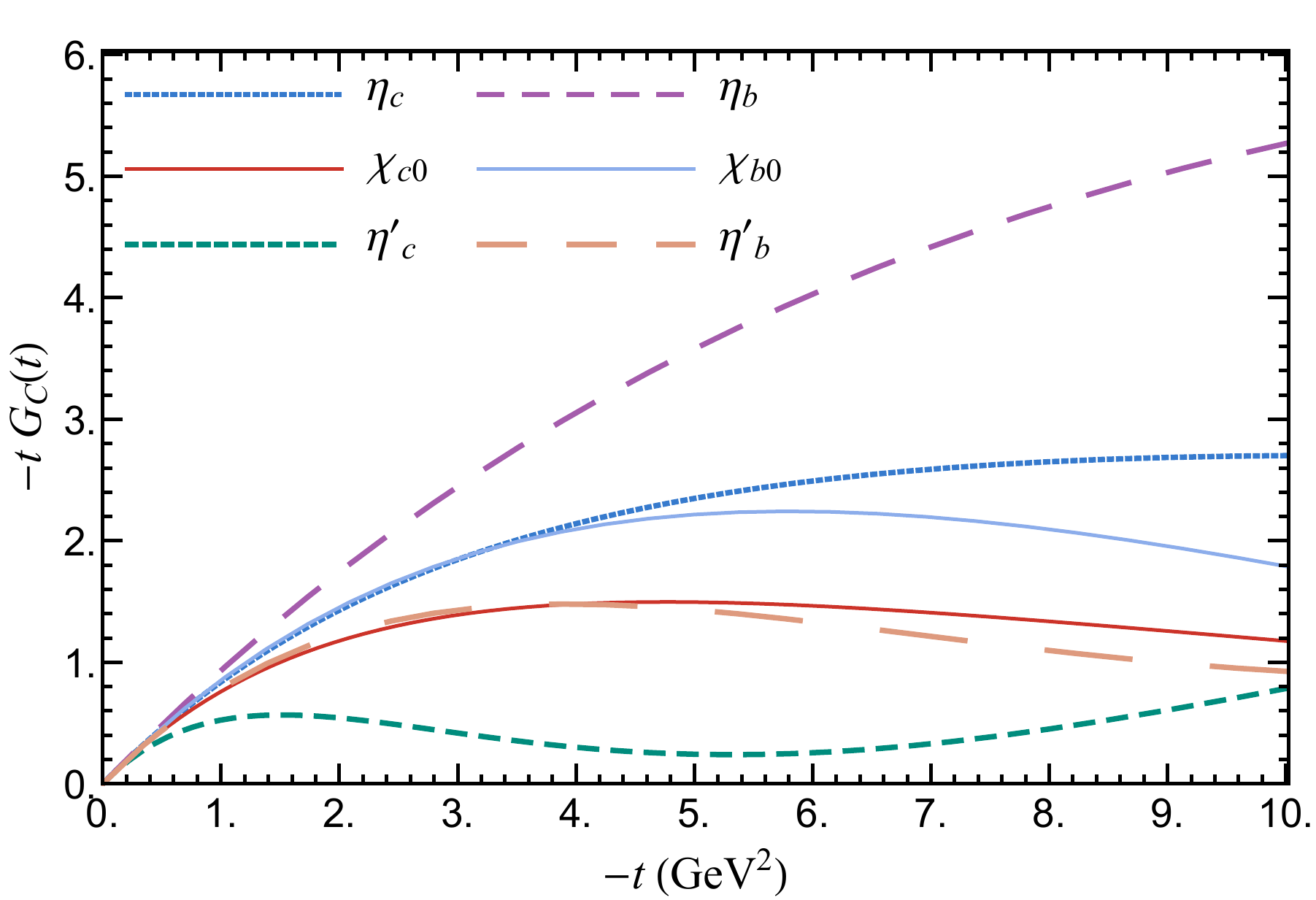}}\\
\end{tabular}
\begin{tabular}{cc}
\subfloat{\includegraphics[scale=0.478]{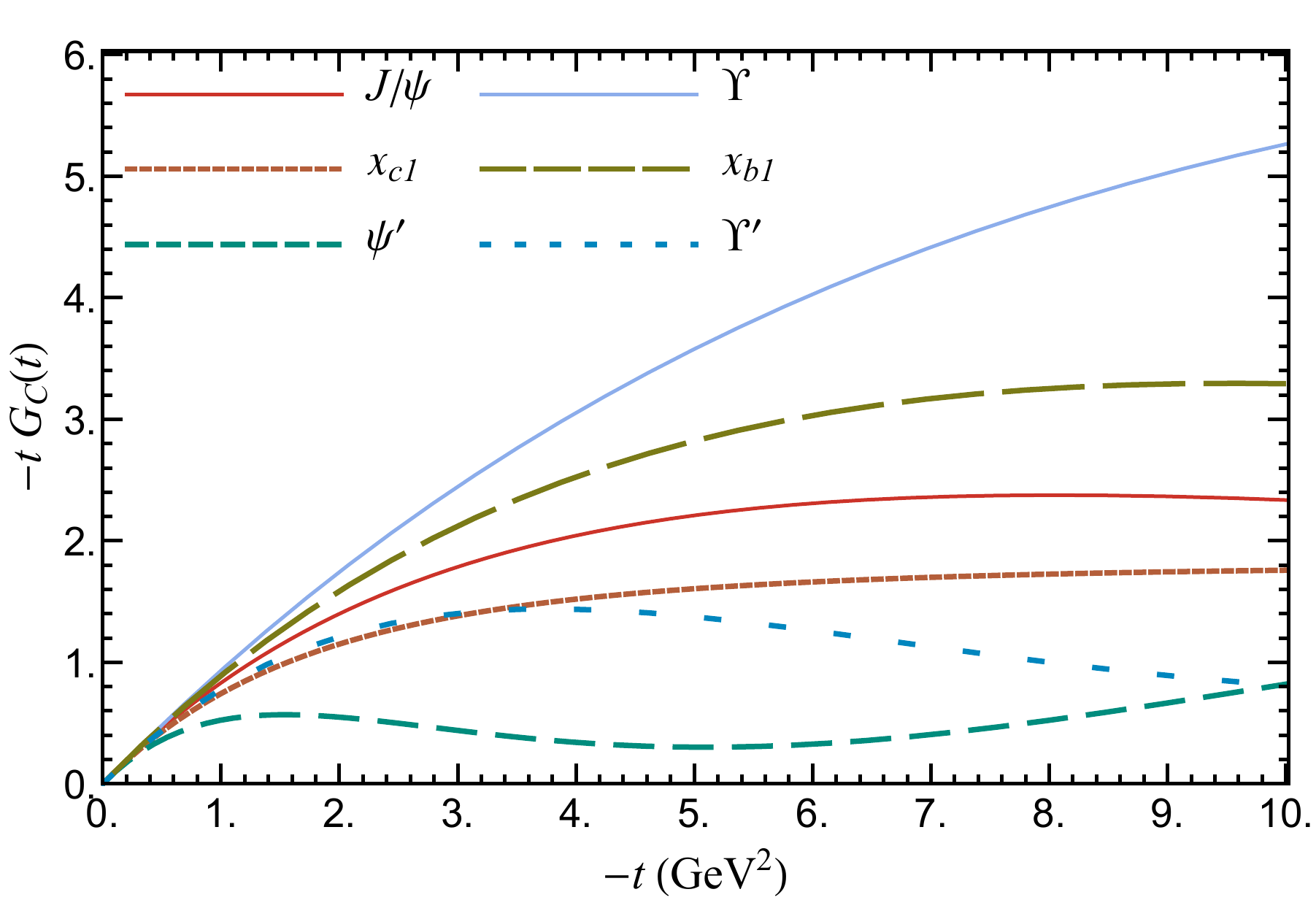}}\\
\end{tabular}
\caption{ $- t\,G_C(t)$ vs $-t$ for  (pseudo) scalar mesons (left panel) and (axial) vector mesons (right panel) in the BLFQ approach.}\label{fig:charge_scalar_vector}
\end{figure*}

\begin{figure}
\begin{tabular}{cc}
\subfloat{\includegraphics[scale=0.485]{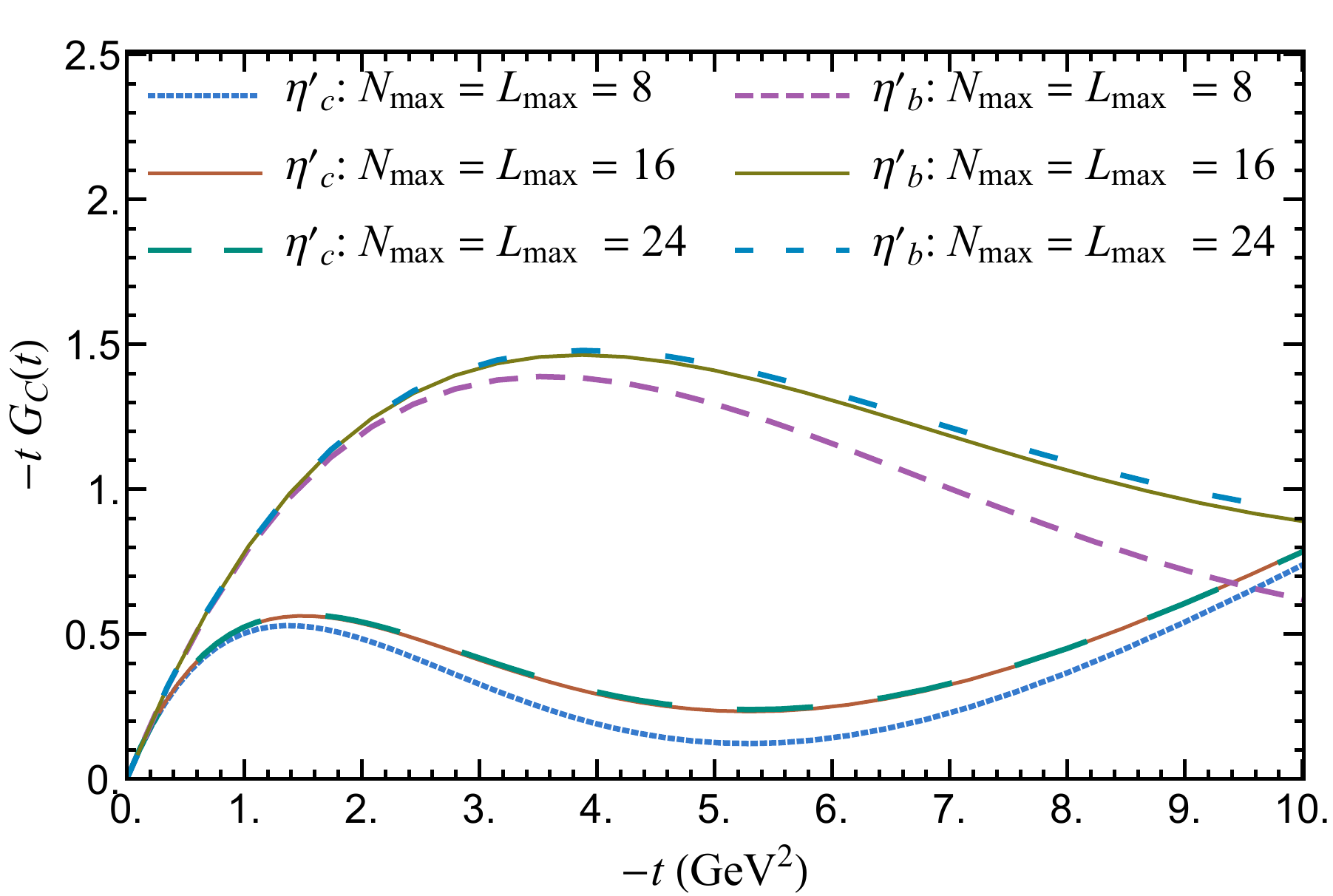}}\\
\end{tabular}
\begin{tabular}{cc}
\subfloat{\includegraphics[scale=0.473]{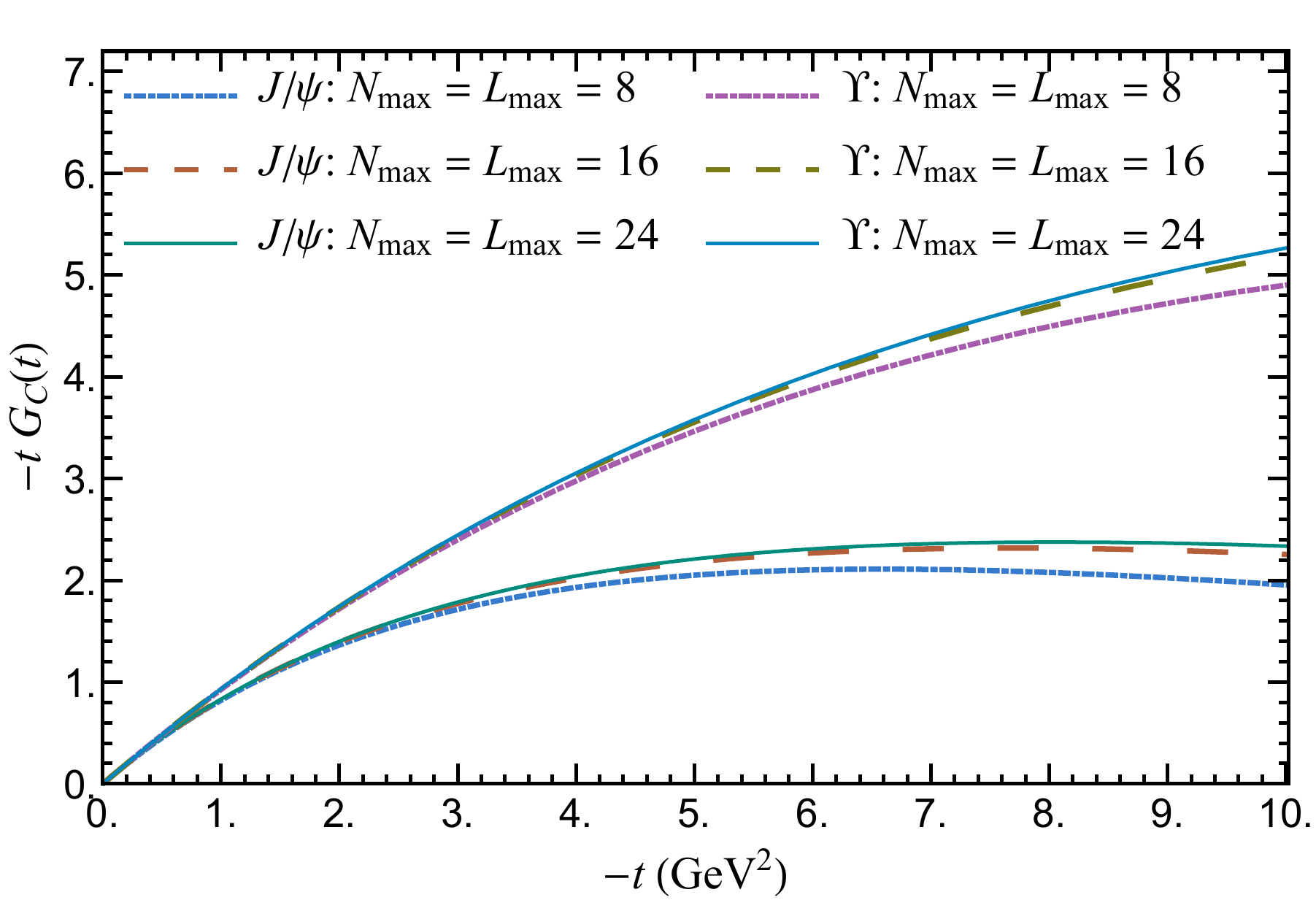}}\\
\end{tabular}
\caption{ $-t G_C(t)$ vs $-t$  for  $\eta^\prime_c$ and $\eta^\prime_b$ (left panel) and $J/\psi$ and $\Upsilon$ (right panel) with different $N_{\text{max}}= L_{\text{max}}$  in the BLFQ approach.}\label{fig:convergence}
\end{figure}
 \begin{table}
\caption{The charge mean squared radii $\langle r^2 \rangle$ of (pseudo) scalar charmonia and bottomonia [Eq.~\eqref{eq:charge_radius}] with $N_{\text{max}} = L_{\text{max}} =24$.  The difference between the $N_{\text{max}} = L_{\text{max}} = 24$ and 8 values are presented as the uncertainty for the BLFQ results. We compare our results with those of the Contact Interaction (CI), Lattice and Dyson-Schwinger Equation (DSE) methods.
}\label{tab:radii_scalar}
 \centering
\begin{tabular}{ccc ccc ccc c}

\toprule
($\text{fm}^2$) & $\eta_c$  & $\chi_{c0}$ & $\eta'_c$ &  $\eta_b$  & $\chi_{b0}$ & $\eta'_b$    \\

\colrule

\multirow{1}{*}{this work (BLFQ)} & \multirow{1}{*}{0.043(5)} & 0.07(1) & 0.149(8) & 0.016(1) & 0.037(1) & 0.056(2) &  \\
\multirow{1}{*}{this work (SBL)} & \multirow{1}{*}{0.073} &0.145  &0.218 &0.0295  &0.0591  &0.0886  &  \\
\multirow{1}{*}{CI} \cite{Bedolla:2015mpa,Raya:2017ggu,Bedolla:2016yxq} & \multirow{1}{*}{0.044} &  &  &0.012 & & &  \\
\multirow{1}{*}{Lattice \cite{Dudek:2006ej}} & \multirow{1}{*}{0.063(1)} & 0.095(6) &    \\

\multirow{1}{*}{DSE} \cite{Maris:2006ea,Bhagwat:2006pu} & \multirow{1}{*}{0.048(4)} &  &  & & & &  \\

\botrule
\end{tabular}
\end{table}
\subsection{The EM FFs and the associated static moments}
In this subsection, we present results for the EM FFs and the associated static moments. We start by presenting the charge FFs $G_C(t)$ for  (pseudo) scalar and  (axial) vector mesons in Fig.~\ref{fig:charge_scalar_vector}. Note for the (pseudo) scalar mesons of the left panel in Fig.~\ref{fig:charge_scalar_vector}, $\eta_c$, $\chi_{c0}$, $\eta^\prime_c$, $\eta_b$, $\chi_{b0}$ and $\eta^\prime_b$, Eq.~\eqref{eq:ff_LFWF} directly produces the charge FFs as $G_C(t)\equiv I_{0, 0} (t)$, whereas for the vector mesons of the right panel in Fig.~\ref{fig:charge_scalar_vector}, $J/\psi$, $\chi_{c1}$, $\psi^\prime$, $\Upsilon$, $\chi_{b1}$ and $\Upsilon^\prime$, the GK prescription [Eq.~\eqref{eq:c_amplitude}] is used to calculate the charge FFs. The FFs for the radially excited charmonia, $\eta^{\prime}_c$ and $\psi^\prime$ , exhibit a tendency to develop a node while the corresponding states in bottomonium show this tendency only at larger values of $-t$ (not shown). Nodes in FFs are common features for excited states in non-relativistic systems.

We present the charge FF results for four selected mesons,~$\eta_c^\prime$,~$\eta_b^\prime$,~$J/\psi$ and~$\Upsilon^\prime$ at a sequence of $N_{\text{max}} = L_{\text{max}} = 8,~16,$ and $24$ values to gain a perspective on their convergence. On the left panel of Fig.~\ref{fig:convergence}, we present the convergence of $-t\, G_C(t)$ for the  pseudo scalar mesons, and in the right panel, we present the same observable for the vector mesons. The results show a good convergence trend over this range of $-t$ as evident by finding that the $N_{\text{max}} = L_{\text{max}} = 24$ and $N_{\text{max}} = L_{\text{max}} = 16$ results are nearly coincident with each other in contrast with the $N_{\text{max}} = L_{\text{max}}=8$ results presented in Fig.~\ref{fig:convergence}. This observed convergence in the FFs is reassuring since the charmonia and bottomonia spectroscopy are also reasonably well converged at $N_{\text{max}} = L_{\text{max}} = 24$ ~\cite{Li:2015zda}. Therefore, we only present our FF and GPD results calculated  with $N_{\text{max}} = L_{\text{max}} = 24$. The difference between the $N_{\text{max}} = L_{\text{max}} = 24$ and 8 values are presented as our uncertainty estimate.

\begin{table}
\caption{The charge mean squared radii $\langle r^2 \rangle$ [Eq.~\eqref{eq:charge_radius}] for (axial) vector charmonia and bottomonia. The difference between the $N_{\text{max}} = L_{\text{max}} = 24$ and 8 values are presented as the uncertainty for the BLFQ results.
}\label{tab:radii_vector}
 \centering
\begin{tabular}{ccc ccc ccc c}

\toprule
($\text{fm}^2$)& $J/\psi$ & $\chi_{c1}$ & $\psi^\prime$ & $\Upsilon$ &
$\chi_{b1}$ &
$\Upsilon^\prime$  \\

\colrule
\multirow{1}{*}{this work (BLFQ)} & \multirow{1}{*}{0.045(3)} & 0.075(2) & 0.15(1) & 0.016(1) & 0.0270(4) & 0.057(3) &  \\
\multirow{1}{*}{this work (SBL)} & \multirow{1}{*}{0.077} & 0.081&0.221 & 0.02996 & 0.0315  &0.08899  &  \\
\multirow{1}{*}{CI} \cite{Raya:2017ggu} & \multirow{1}{*}{0.068} &  &  &0.038 & & &  \\
\multirow{1}{*}{Lattice \cite{Dudek:2006ej}} & \multirow{1}{*}{0.066(2)} &     \\

\multirow{1}{*}{DSE} \cite{Maris:2006ea,Bhagwat:2006pu} & \multirow{1}{*}{0.052(3)} &   \\

\botrule
\end{tabular}
\end{table}

\begin{figure*}
\begin{tabular}{cc}
\subfloat{\includegraphics[scale=0.48]{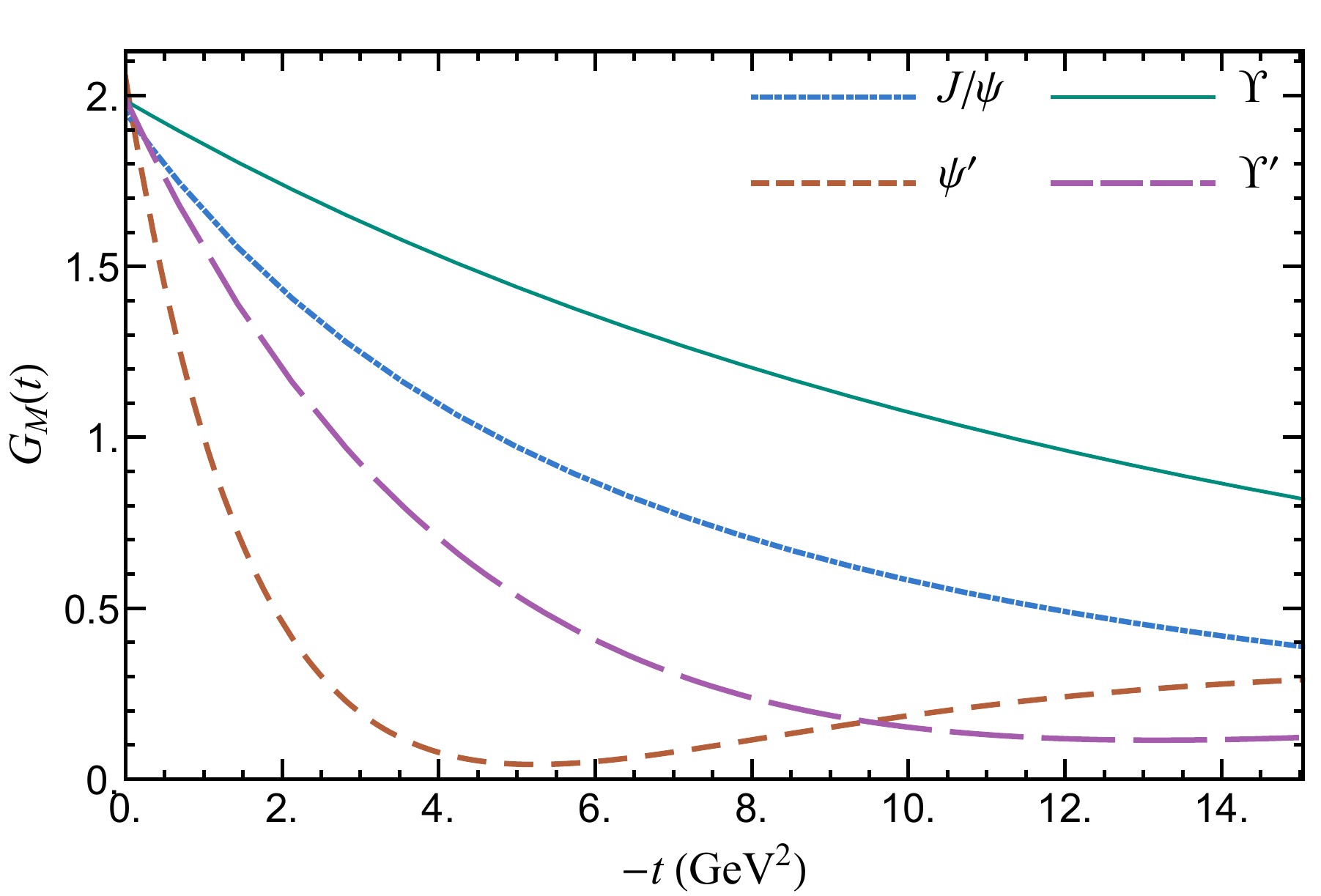}}\\
\end{tabular}
\begin{tabular}{cc}
\subfloat{\includegraphics[scale=0.475]{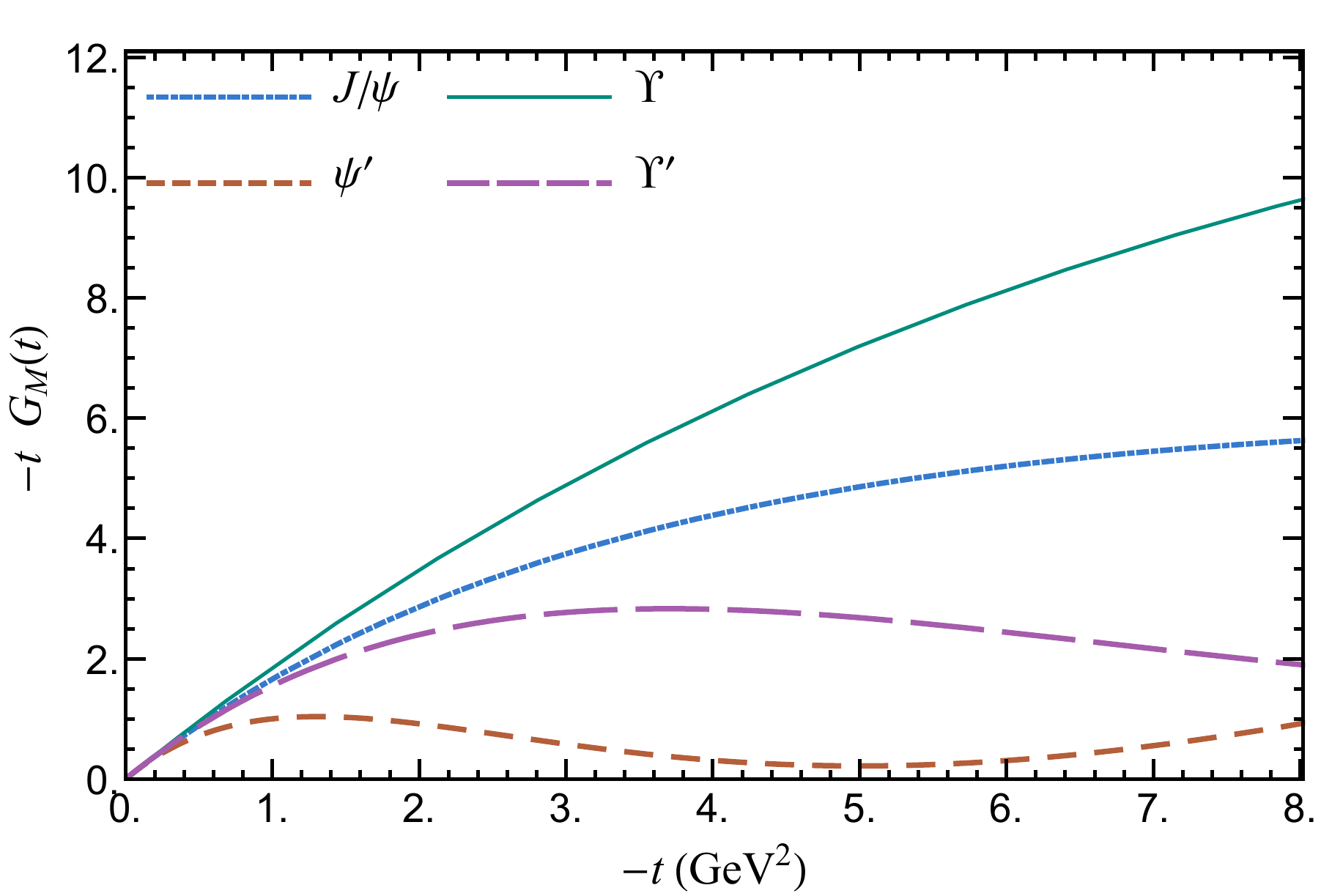}}\\
\end{tabular}
\caption{ The magnetic FFs $G_M(t)$ [Eq.~\eqref{eq:m_amplitude}] for selected vector charmonia and bottomonia states in the BLFQ approach.}\label{fig:magnetic_formfactor}
\end{figure*}

We now turn our attention to the charge mean squared radii of charmonia and bottomonia calculated both in the BLFQ and the SBL approaches. We note again that the charge mean squared radius is an artificial quantity defined with the neglect of the contribution of the anti-quark to the form factor. Table~\ref{tab:radii_scalar} lists the charge mean squared radii (in $\text{fm}^2$) of selected (pseudo) scalar mesons, and Table~\ref{tab:radii_vector} lists those of selected (axial) vector mesons. We see from Tables~\ref{tab:radii_scalar} and \ref{tab:radii_vector} that the charge radii of the selected charmonia states  are larger than that of their counterparts in bottomonia. This relative relationship is found for both BLFQ and SBL results as well as for the available CI results. This observation about the relative radii  can be understood simply from the tendency towards the non-relativistic limit with increasing quark mass. It is also noted from Table~\ref{tab:radii_scalar} that the charge radius of $\eta_c$ is smaller than that of $\chi_{c0}$ both in the BLFQ and SBL approaches, and this relationship is consistent with the Lattice results~\cite{Dudek:2006ej}.

We note Tables~\ref{tab:radii_scalar} and \ref{tab:radii_vector} show significant differences among the results calculated in different formalisms which is reasonable considering the major distinctions among the formalisms. For example, in Ref.~\cite{Maris:2006ea}, the  DSE results were calculated describing $J/\psi$ by the solutions of the homogeneous Bethe-Salpeter equations (BSE) in rainbow-ladder truncation. The DSE results also reflect the adoption of an effective running coupling via one-gluon exchange. Among the many differences with our BLFQ results we note our use of a fixed coupling. Furthermore, in Refs.~\cite{Bedolla:2015mpa,Raya:2017ggu,Bedolla:2016yxq}, the CI results were calculated using contact interactions within the framework of the DSE and BSE. Despite several differences between the CI and BLFQ approaches, there is however a reasonable agreement among the resulting charge radii for the mesons $\eta_c$ and $\eta_b$. We also observe that for each meson the radius calculated in the SBL approach is larger than the radius calculated in the BLFQ approach. This observation can be understood from the fact that SBL results are  produced by only taking the leading basis function into account, which means that the radius is controlled by the dominant mode and by the confining length scale, while the BLFQ includes the gluon exchange, an attractive interaction.

\begin{figure*}
\includegraphics[scale=0.48]{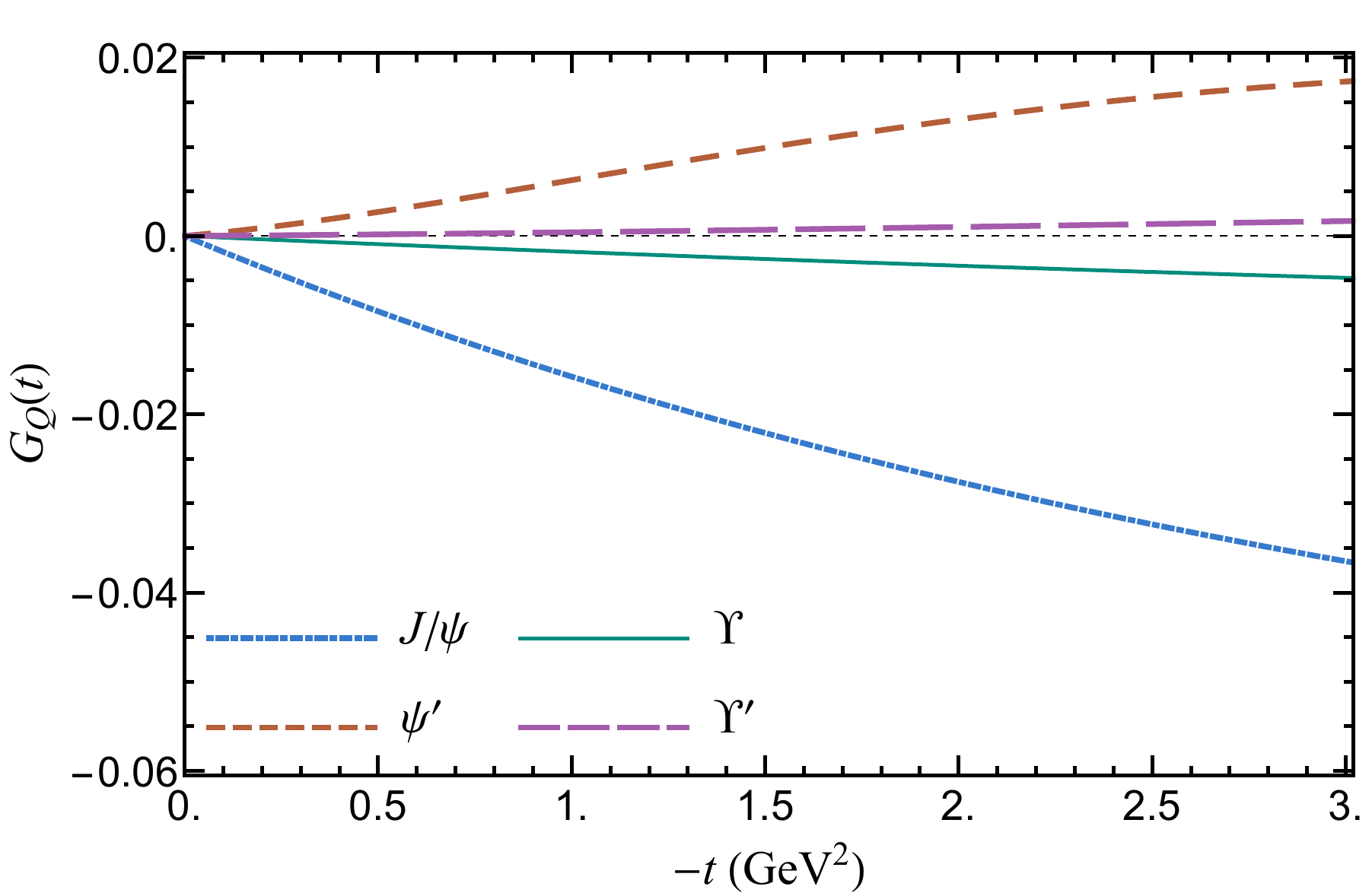}
\caption{ The quadrupole FFs $G_Q(t)$ [Eq.~\eqref{eq:q_amplitude}] for selected vector mesons in the BLFQ approach.} \label{fig:quadrupole_formfactor}
\end{figure*}

\begin{figure}
\begin{tabular}{cc}
\subfloat{\includegraphics[scale=0.469]{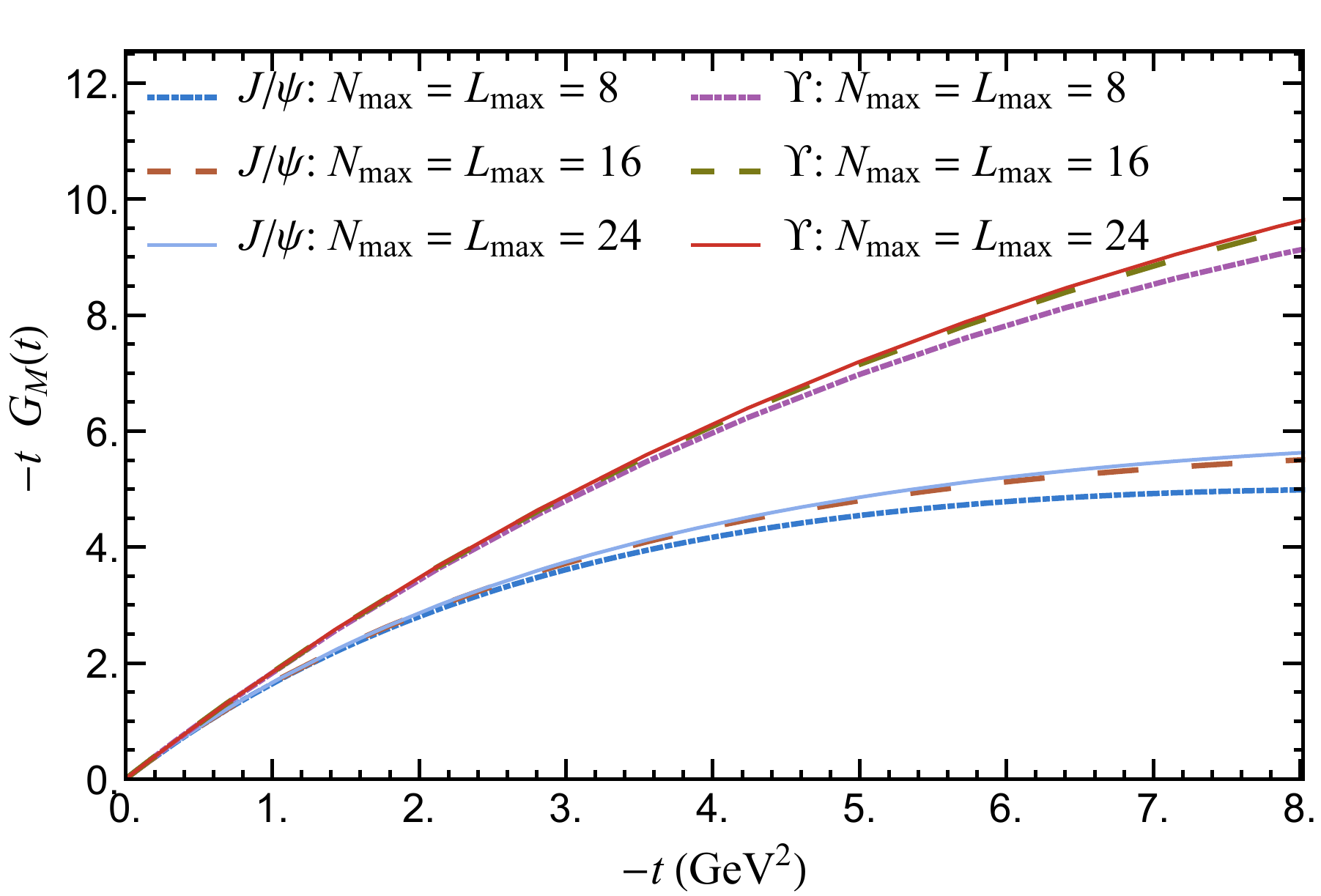}}\\
\end{tabular}
\begin{tabular}{cc}
\subfloat{\includegraphics[scale=0.489]{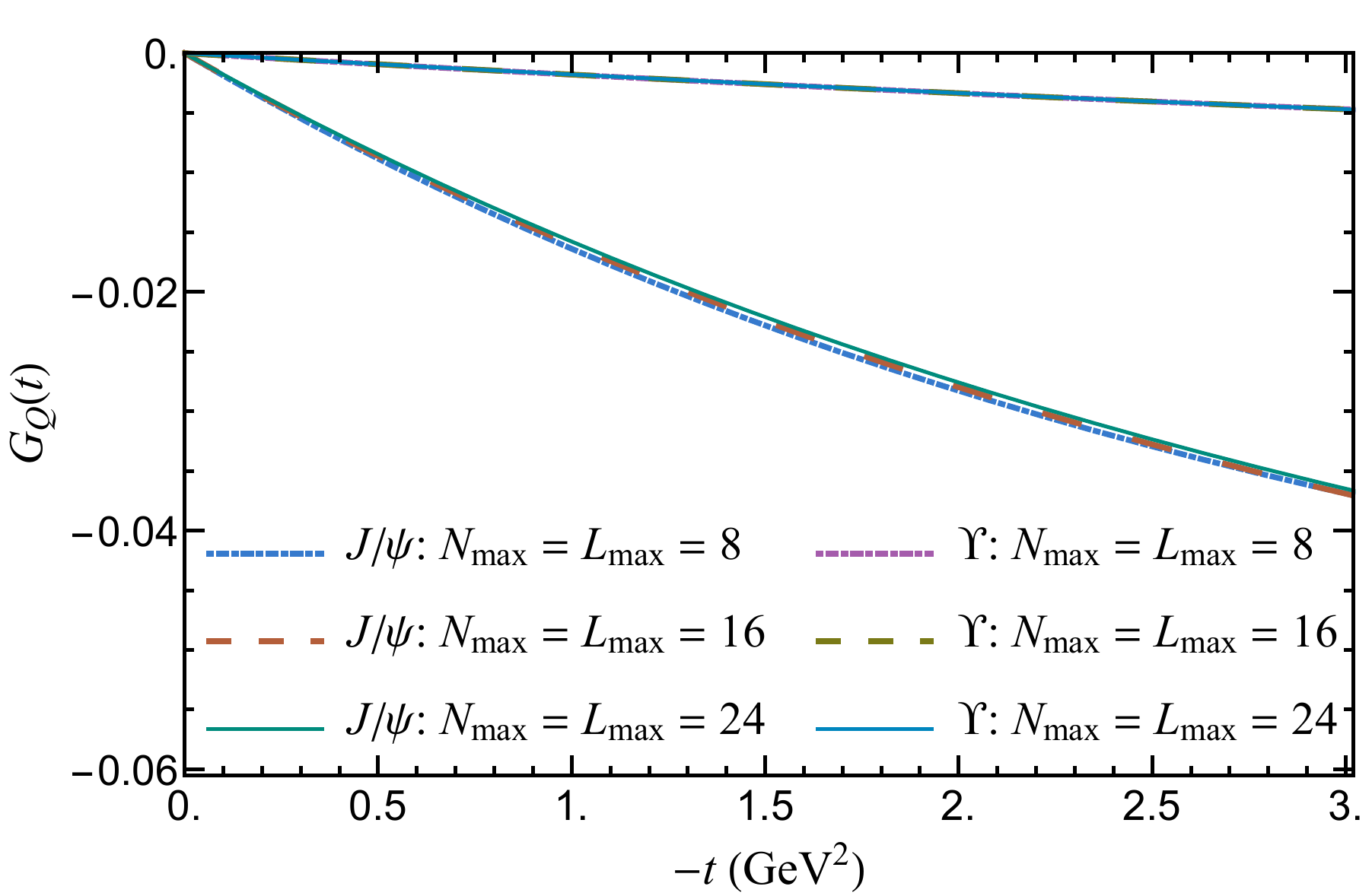}}\\
\end{tabular}
\caption{The comparison of magnetic FFs  $G_M(t)$ (left panel)  and the quadrupole FFs $G_Q(t)$ (right panel) for $1^3S_1$ ($J/\psi$ and $\Upsilon$) with different $N_{\text{max}}= L_{\text{max}}$ in the BLFQ approach.} \label{fig:convergence_1}
\end{figure}

Next, we present the magnetic FFs $G_M(t)$  and the quadrupole FFs $G_Q(t)$ of vector mesons calculated with $N_{\text{max}} = L_{\text{max}} = 24$ in the BLFQ approach. Figure~\ref{fig:magnetic_formfactor} presents the magnetic FFs $G_M(t)$ [Eq.~\eqref{eq:m_amplitude}] and Figure~\ref{fig:quadrupole_formfactor} presents the quadrupole FFs $G_Q(t)$ [Eq.~\eqref{eq:q_amplitude}]. As we presented the convergence of the charge FFs with respect to basis truncation above, we present in Fig.~\ref{fig:convergence_1} the convergence of the $-t\, G_M(t)$ (left panel) and $G_Q(t)$ (right panel) with respect to $N_{\text{max}} =L_{\text{max}}$. The results, again, show a good convergence trend since  the $N_{\text{max}} = L_{\text{max}} = 24$ and $N_{\text{max}} = L_{\text{max}} = 16$ values are in close agreement over the range of $-t$ presented.  On the other hand these same form factors have visibly larger differences from the results at $N_{\text{max}} = L_{\text{max}} = 8$.

The magnetic and quadrupole moments associated with the vector mesons ~$J/\psi$, $\psi^\prime$, $\Upsilon$ and $\Upsilon^{\prime}$ are calculated and presented in Tables~\ref{tab:magnetic_moment} and ~\ref{tab:quadrupole_moment}, respectively. The magnetic and quadrupole moments calculated in the SBL approach are 2.0 and -1.0, respectively, the canonical values, as expected. The SBL results can also be understood  by analyzing the helicity amplitudes $I_{+,0}$ and $I_{+,-}$ in Eqs.~\eqref{eq:m_amplitude},~\eqref{eq:q_amplitude},~\eqref{eq:magnetic_moment} and ~\eqref{eq:quadrupole_moment}. In the SBL approach, there is no contribution from either of these amplitudes to the magnetic and quadrupole moments, and that is because only the leading basis function contribution(s) of the LFWFs is (are) taken into account. The BLFQ magnetic moments for the mesons ~$J/\psi$, $\Upsilon$ and $\Upsilon^{\prime}$ are below 2.0 while the results from the cited literature are above 2.0 where available. This led us to make additional checks of our calculations to confirm the accuracy of our results. It is interesting to note that theoretical results for the rho meson are often below 2.0 as well ~\cite{Shtokhamer:1973ze,Samsonov:2002mz,Jaus:2002sv,Choi:2004ww}. For example, in Ref.~\cite{Jaus:2002sv}, the investigation in the framework of a covariant extension of the LF formalism has found the rho meson magnetic moment to be 1.83. Another investigation in the LF quark model~\cite{Choi:2004ww}  has found it to be 1.92 and an investigation in the framework of QCD sum rules~\cite{Samsonov:2002mz} has found it to be 1.5$\pm0.3$.

Inspecting our results in Table~\ref{tab:magnetic_moment}, we comment that the magnetic moments of the vector mesons calculated in the BLFQ approach are  closer to corresponding SBL quantities for the case of bottomonia  than for the case of charmonia suggesting that, for this quantity, the role of the gluon exchange interaction is reduced in bottomonium relative to charmonium. Turning to Table~\ref{tab:quadrupole_moment}, we find that that the quadrupole moment result  for $J/\psi$  calculated in the BLFQ approach is closer to the corresponding CI result than  to the DSE and Lattice results. The magnetic and quadrupole moments calculated in the BLFQ approach clearly show the deviations from corresponding SBL results ($\mu=2.0$ and $\text{Q} =-1.00$ ) which simply underscores the fact that deviations from SBL values point to the gluon exchange dynamics within heavy quarkonia.

\begin{table}
\caption{Magnetic moments $\mu$ [Eq.~\eqref{eq:magnetic_moment}] for vector mesons.  The difference between the $N_{\text{max}} = L_{\text{max}} = 24$ and 8 values are presented as the uncertainty for the BLFQ results.
}\label{tab:magnetic_moment}
 \centering
\begin{tabular}{ccc ccc ccc c}

\toprule
  & $ J/\psi$ & $\psi^\prime$ & $ \Upsilon$ & $\Upsilon^\prime$  \\

\colrule

\multirow{1}{*}{this work (BLFQ)} & \multirow{1}{*}{1.952(3)} & 2.05(2) & 1.985(1) & 1.992(1) &  \\
\multirow{1}{*}{this work (SBL)} & \multirow{1}{*}{2.00} & 2.00 & 2.00 & 2.00 &  \\
\multirow{1}{*}{CI} \cite{Raya:2017ggu} & \multirow{1}{*}{2.047} &  & 2.012 & &  \\
\multirow{1}{*}{Lattice \cite{Dudek:2006ej}} & \multirow{1}{*}{2.10(3)}  &    \\

\multirow{1}{*}{DSE} \cite{Maris:2006ea,Bhagwat:2006pu} & \multirow{1}{*}{2.13(4)} &  \\

\botrule
\end{tabular}
\end{table}

\begin{table}
\caption{Quadrupole moments  $(\text{Q}\times M^2 )$ [Eq.~\eqref{eq:quadrupole_moment}] for vector mesons. The results are presented as unitless, and the difference between the $N_{\text{max}} = L_{\text{max}} = 24$ and 8 values are presented as the uncertainty for the BLFQ results.
}\label{tab:quadrupole_moment}
 \centering
\begin{tabular}{ccc ccc ccc c}

\toprule
 & $J/\psi$  & $\psi^\prime$ & $\Upsilon$ &
$\Upsilon^\prime$  \\

\colrule

\multirow{1}{*}{this work (BLFQ)} & \multirow{1}{*}{-0.78(2)}  & 0.2(2) & -0.731(9)& 0.1(1)&\\
\multirow{1}{*}{this work(SBL)} & \multirow{1}{*}{-1.000}  & -1.000 & -1.000& -1.000&\\
\multirow{1}{*}{CI} \cite{Raya:2017ggu} & \multirow{1}{*}{-0.748} &  & -0.704 & &  \\
\multirow{1}{*}{Lattice \cite{Dudek:2006ej}} & \multirow{1}{*}{-0.23(2)}  &    \\

\multirow{1}{*}{DSE} \cite{Maris:2006ea,Bhagwat:2006pu} & \multirow{1}{*}{-0.28(1)} &  \\

\botrule
\end{tabular}
\end{table}

\begin{figure*}
\begin{tabular}{cc}
\subfloat[$J/\psi$]{\includegraphics[scale=0.475]{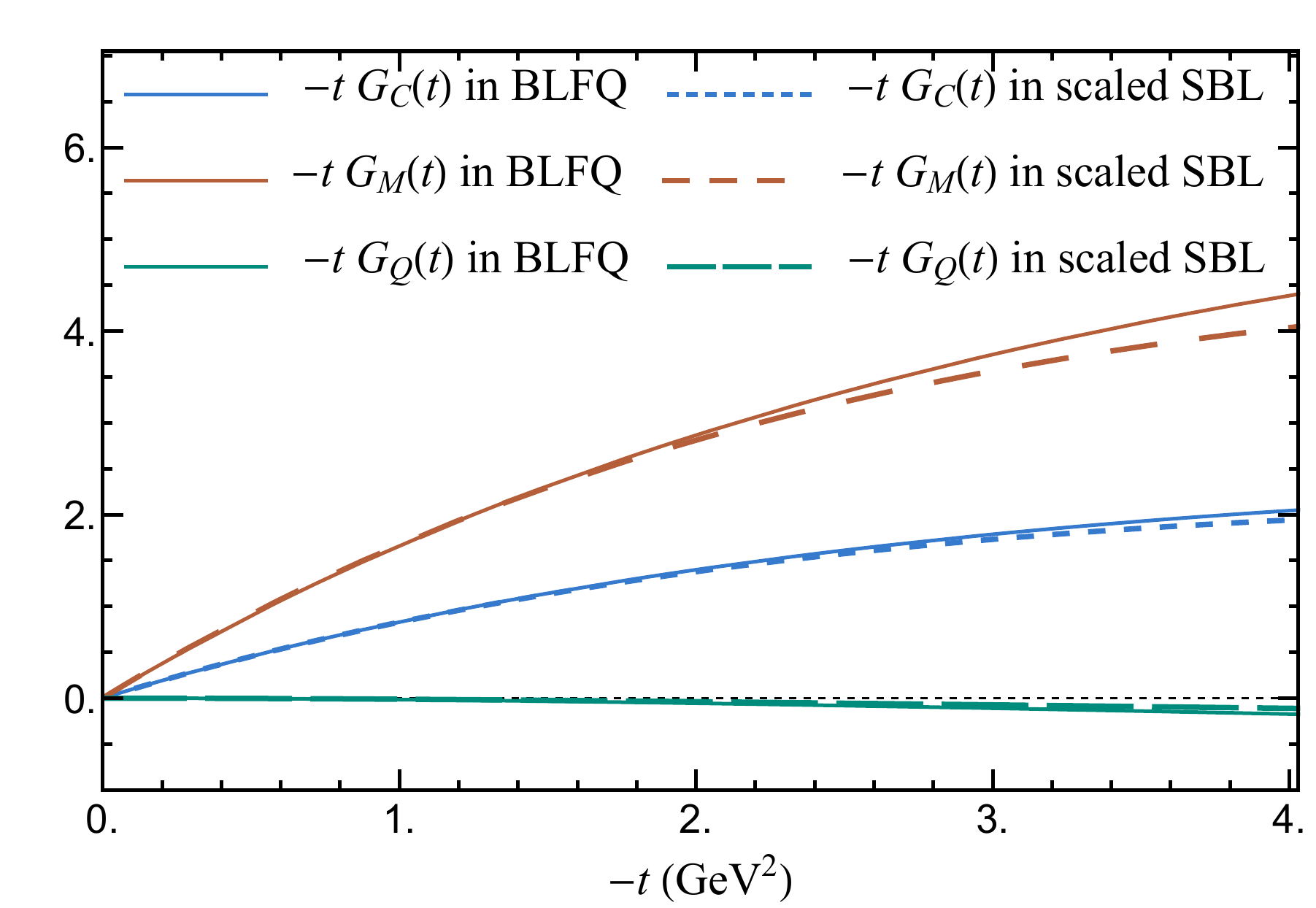}}\\
\end{tabular}
\begin{tabular}{cc}
\subfloat[$\Upsilon$]{\includegraphics[scale=0.483]{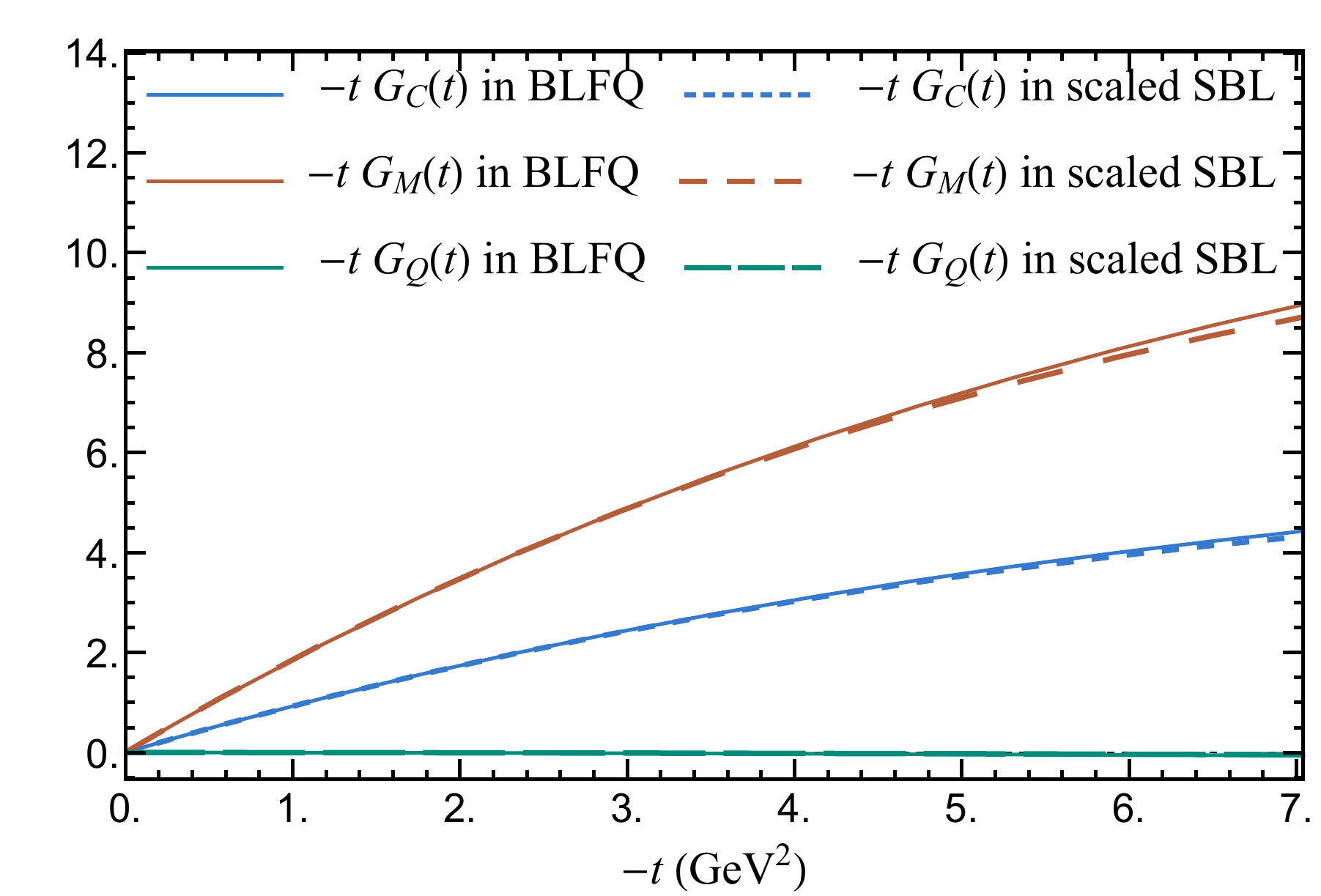}}\\
\end{tabular}
\caption{The comparisons of the EM FFs for $1^3S_1$ [$J/\psi$ (left panel) and $\Upsilon$ (right right)] in the BLFQ and SBL approaches.}\label{fig:BLFQ-SBL}
\end{figure*}

\begin{figure*}
\begin{tabular}{cc}
\subfloat{\includegraphics[scale=0.477]{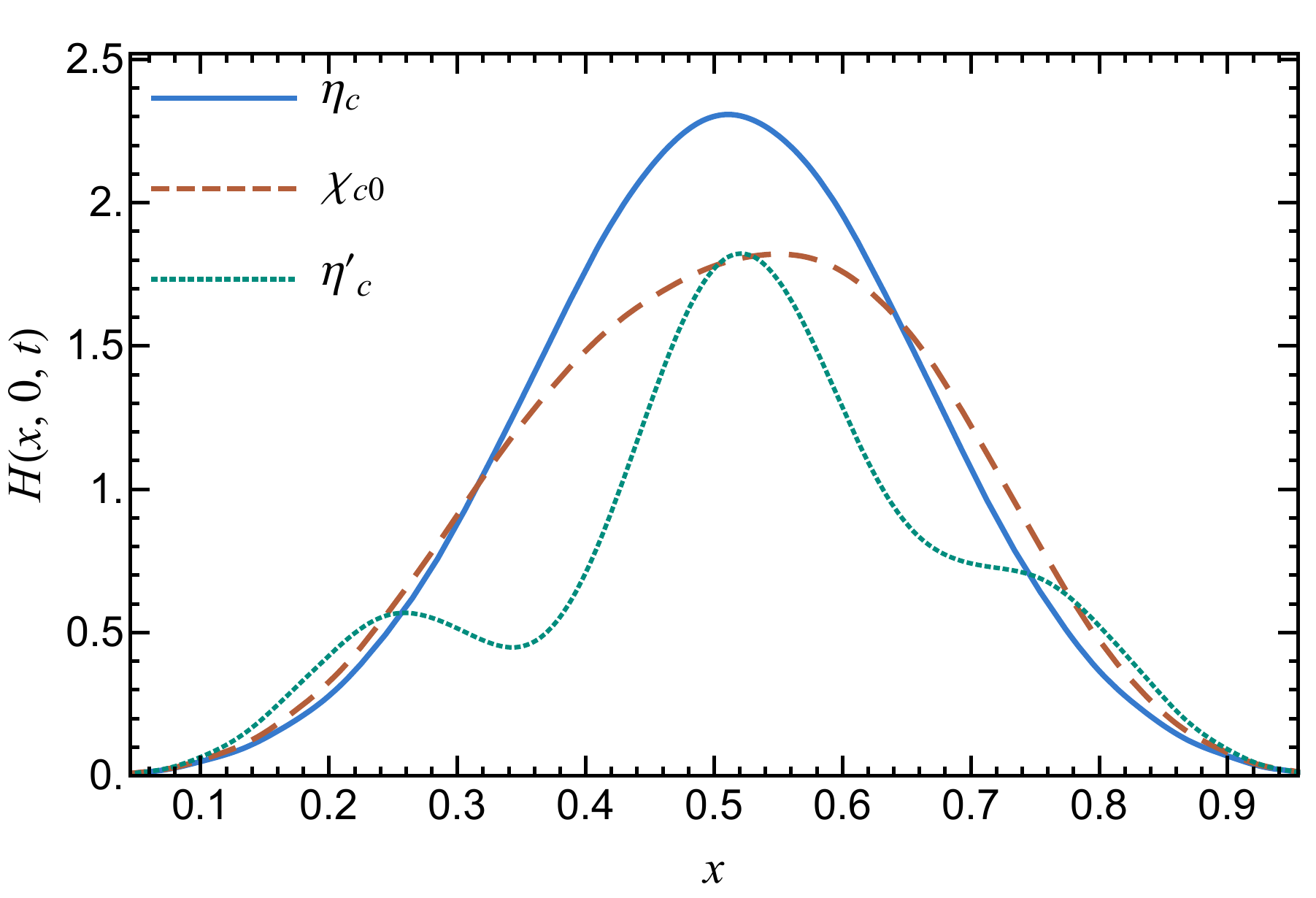}}\\
\end{tabular}
\begin{tabular}{cc}
\subfloat{\includegraphics[scale=0.474]{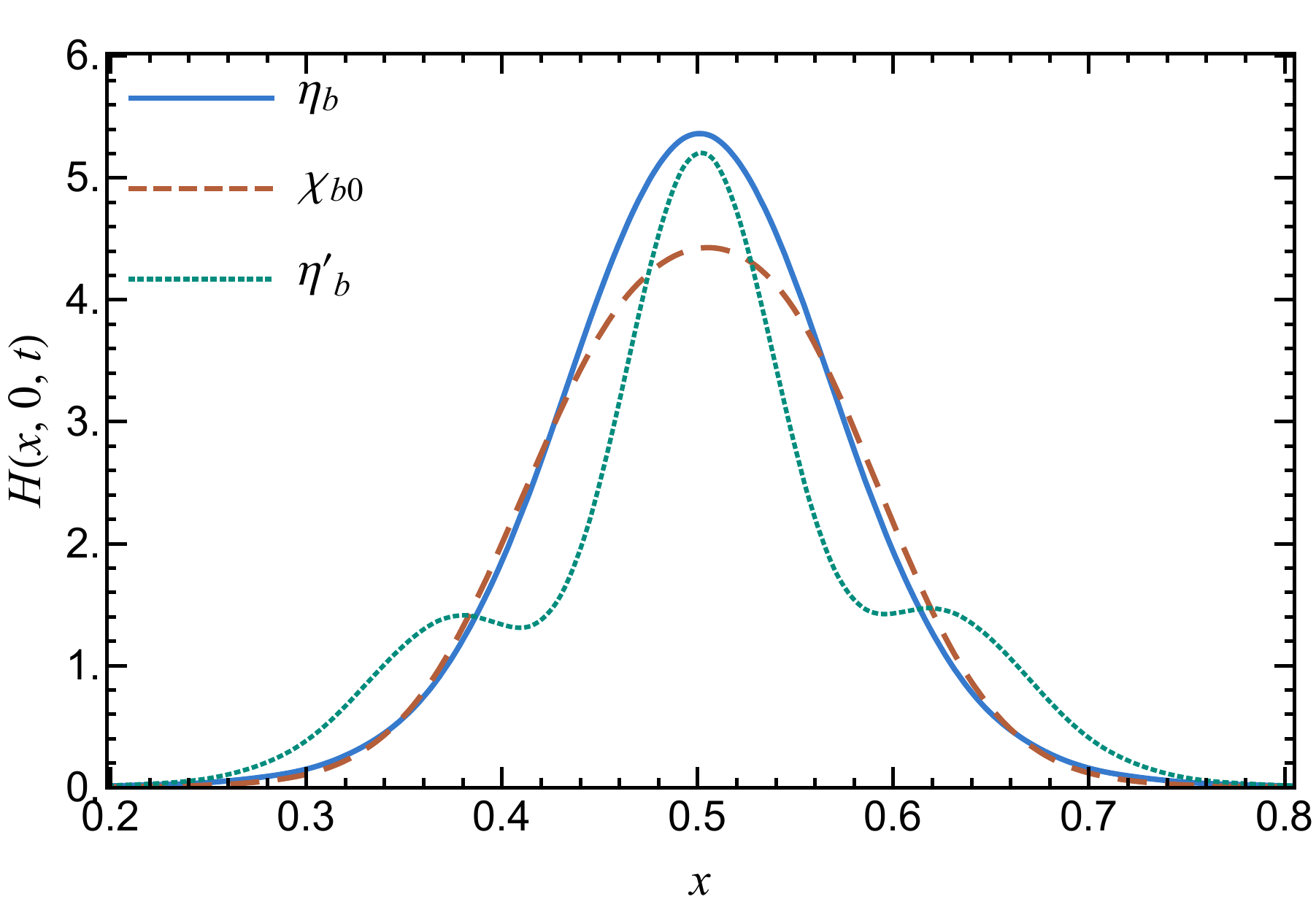}}\\
\end{tabular}
\caption{Helicity non-flip (pseudo) scalar GPDs $H(x, \xi=0, t=-\Delta_\perp^2)$ [Eq.~\eqref{eq:GPDs_LFWF}] for charmonia  (left) and bottomonia (right) at $|t|= 0.765~\text{GeV}^2 $ in the BLFQ approach. Note $t\equiv \Delta^2 = -\Delta_\perp^2$, where $\Delta_\perp$ is the transverse momentum transfer between the initial and final states of the meson and $x$ is the average momentum fraction carried by the quark in the longitudinal direction. \label{fig:gpds_scalar}}
\end{figure*}

Now, before we close this subsection we present comparisons of selected EM FFs calculated in the SBL and BLFQ approaches. However, we first note that there is likely to be a dominant effect from the difference in the  r.m.s. radii between these two approaches. To reduce the impact of this simple difference, we can scale the momentum transfer variable by the appropriate ratio of the charge radii. For this comparison, we select the EM FFs for $J/\psi$ and $\Upsilon$. Following the logic for scaling the momentum transfer for the SBL results, the $-t$ values of the SBL charge FF of a vector meson has been scaled so that its  slope equals to that of corresponding quantity calculated in the BLFQ approach while keeping  both quantities at $t=0$ fixed. Then, the $-t$ values of the SBL magnetic and quadrupole FFs are multiplied by the same factor that sets  the slopes of the SBL and BLFQ charge FFs for the given meson. The scale factor applied to the SBL results for $-t$  is found to be 1.71 for the case of $J/\psi$ and 1.84 for the case of $\Upsilon$. Figure~\ref{fig:BLFQ-SBL} presents the resulting  comparisons of the $J/\psi$ EM FFs (left panel) and the $\Upsilon$ EM FFs (right panel). The BLFQ magnetic and quadrupole FFs in Fig.~\ref{fig:BLFQ-SBL} are very similar  to the corresponding scaled SBL quantities with deviations becoming somewhat visible above approximately $-t = $ 2.4 $~\text{GeV}^2$ in the case of $J/\psi$ and above approximately $-t = $ 6 $~\text{GeV}^2$ in the case of $\Upsilon$. This suggests that the dominant role of gluon exchange dynamics for these form factors is a re-scaling of the size of the system from the size dictated by the confinement scale.

\subsection{Generalized parton distributions}
In this subsection, we present GPDs for a selection of heavy quarkonia starting with the (pseudo) scalar GPDs. For the (pseudo) scalar mesons such as $\eta_c$, $\chi_{c0}$, $\eta^\prime_c$, $\eta_b$, $\chi_{b0}$, and $\eta^\prime_b$, Eq.~\eqref{eq:GPDs_LFWF} directly produces the GPDs, $H(x, \xi=0, t) \equiv V_{0, 0} (x,0, t)$ . In the previous paper~\cite{Adhikari:2016idg}, we have presented 3D plots of (pseudo) scalar GPDs of positronium with the one photon exchange (where the longitudinal confining term in the Hamiltonian is absent, of course). We first present the (pseudo) scalar GPDs at fixed $|t|$ to observe the $x$-dependence of the GPDs in the non-zero momentum transfer limit. Figure.~\ref{fig:gpds_scalar} represents the (pseudo) scalar GPDs of charmonia (left panel) and of their counterpart bottomonia (right panel) at $|t|= 0.765~\text{GeV}^2 $ calculated in the BLFQ approach with $N_{\text{max}}=L_{\text{max}}=24$. It is interesting to observe the change in character of the $x$-dependence of the GPDs between the ground states and the radially excited states, where oscillatory structures emerge.  With our choice of $-t$ value, there is similarity in the structures of corresponding states in charmonium and bottomonium as seen in comparing both panels of Fig.~\ref{fig:gpds_scalar}.

\begin{figure*}
\begin{tabular}{cc}
\subfloat[]{\includegraphics[scale=0.37]{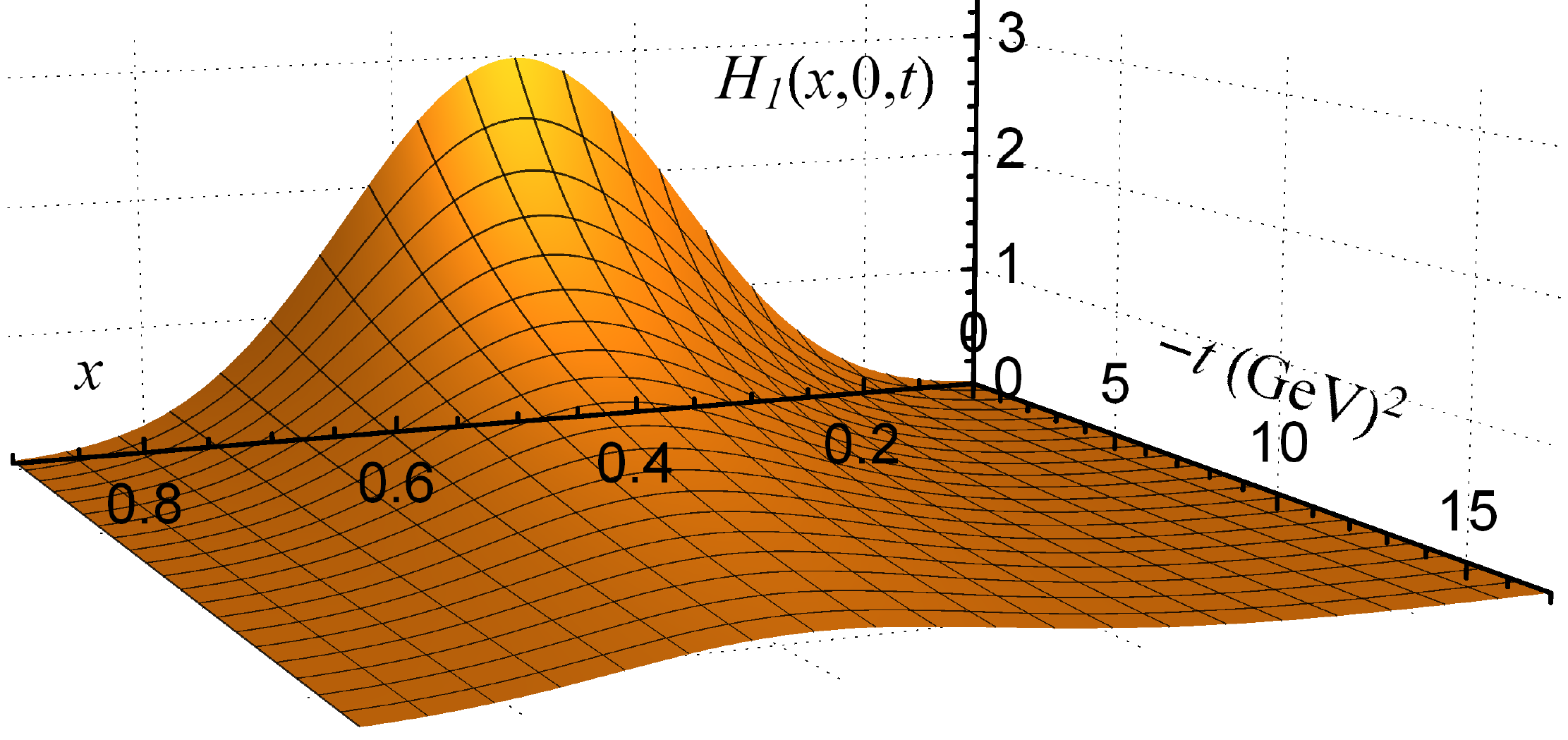}}\\
\subfloat[]{\includegraphics[scale=0.37]{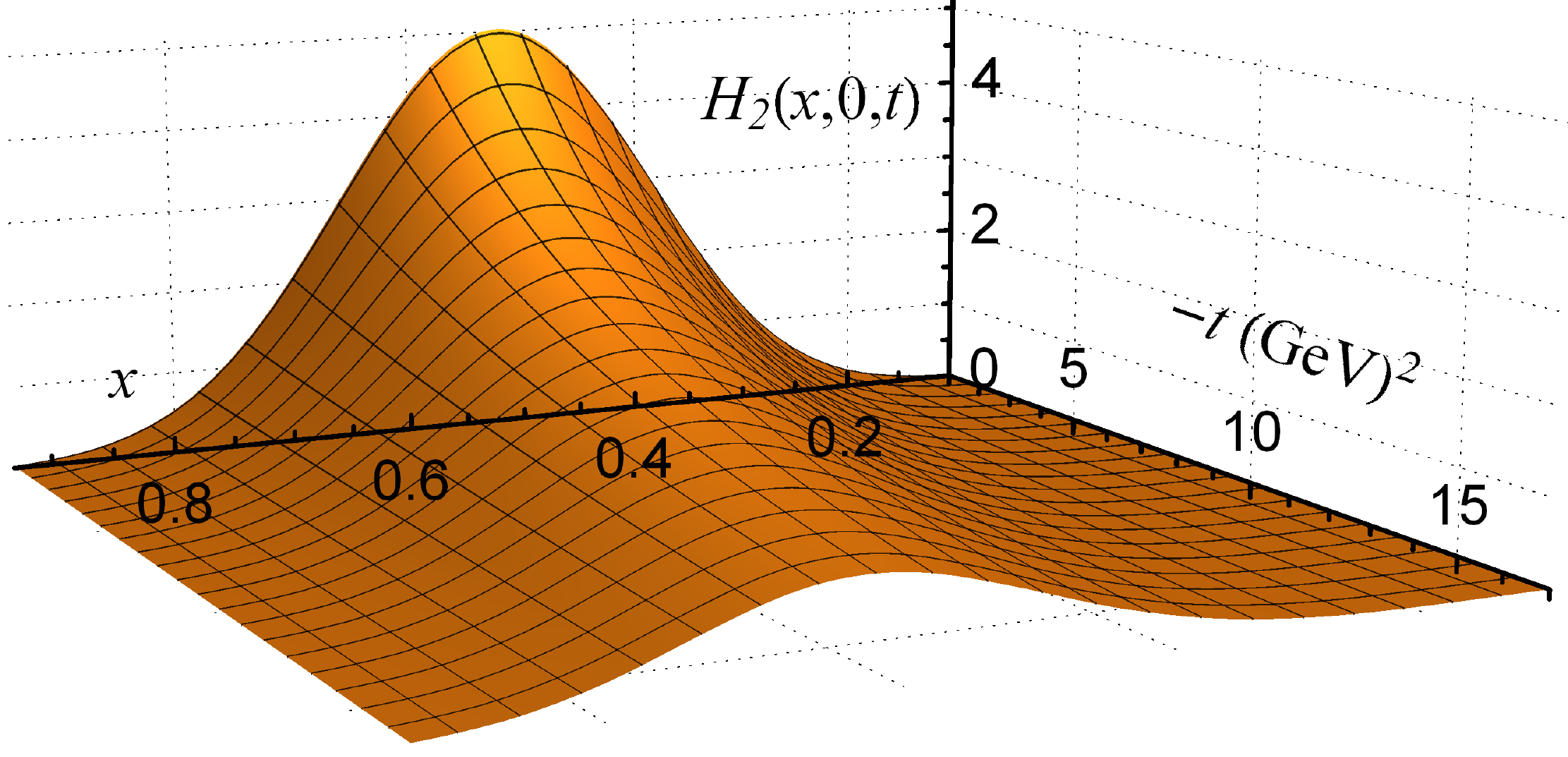}} \\
\end{tabular}
\begin{tabular}{cc}
\subfloat[]{\includegraphics[scale=0.37]{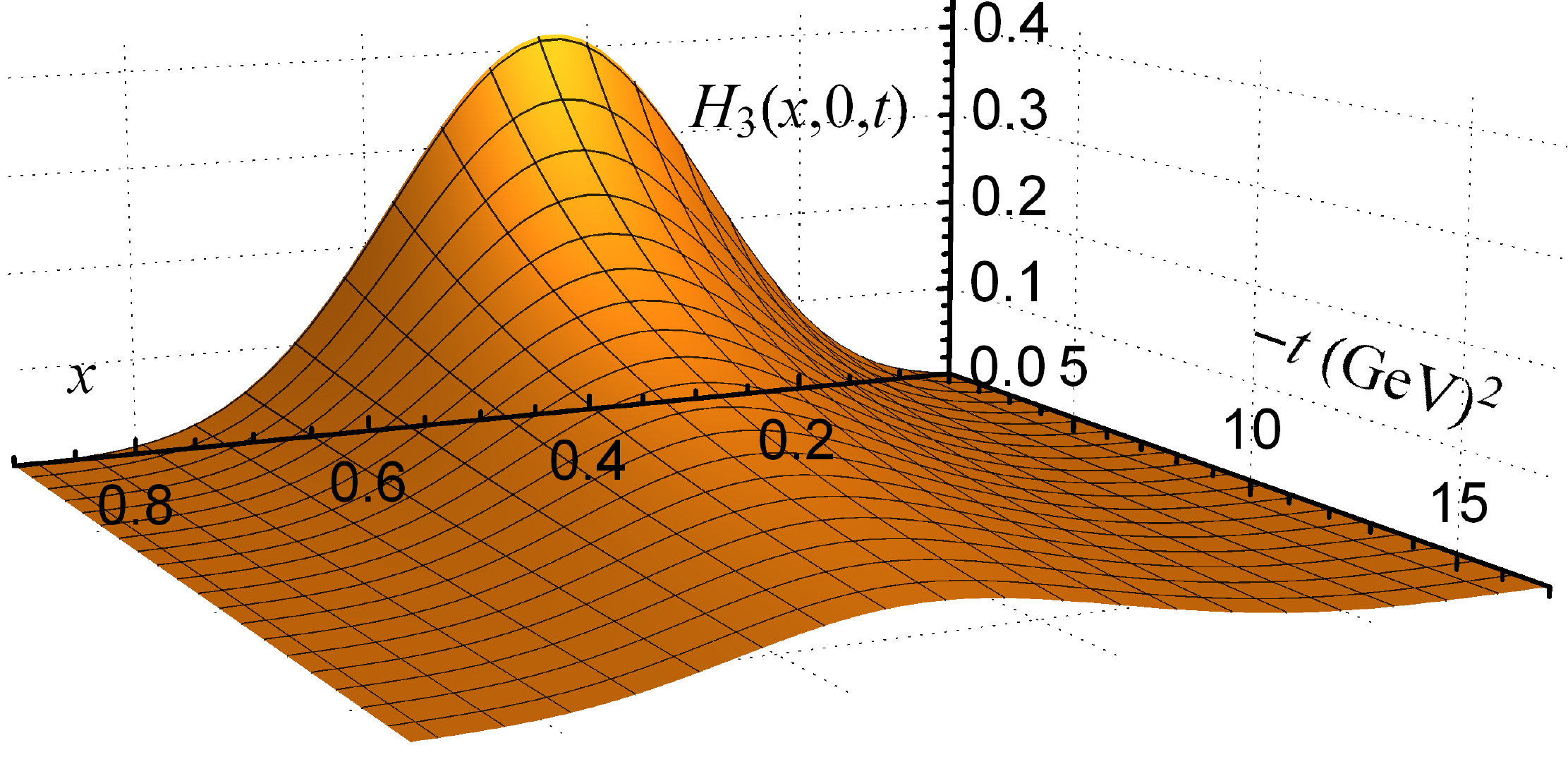}}\\
\subfloat[]{\includegraphics[scale=0.37]{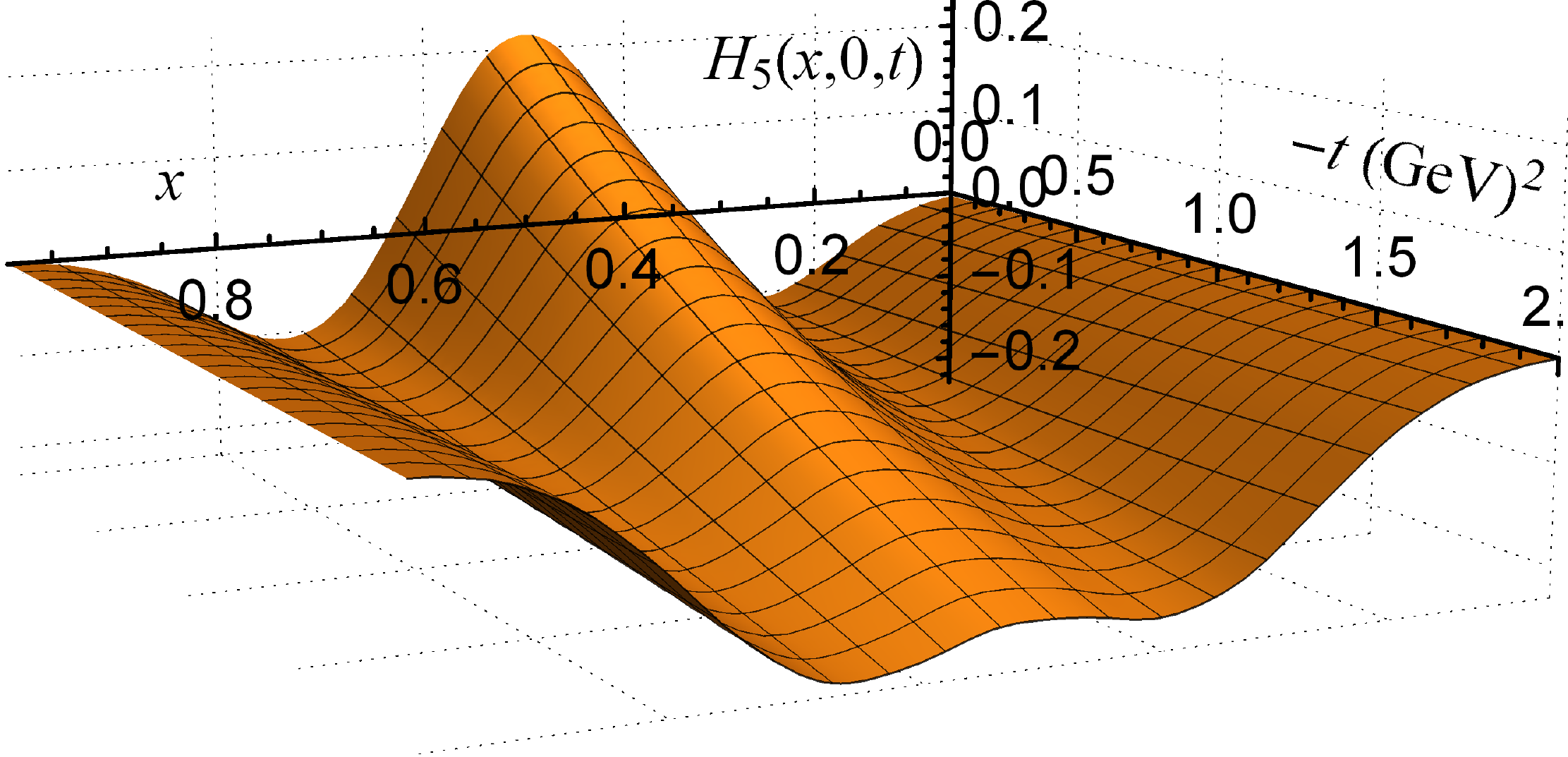}} \\
\end{tabular}
\caption{3D plot of helicity non-flip vector meson GPDs,  $H_i(x, \xi=0, t=-\Delta_\perp^2)$, $i=1, 2, 3, 5$ [Eqs.~\eqref{eq:gpd_amplitude1},~\eqref{eq:gpd_amplitude2},~\eqref{eq:gpd_amplitude3}, and ~\eqref{eq:gpd_amplitude5}] for  $J/\psi \, (1^3S_1)$ in the BLFQ approach. \label{fig:gpds_jpsi}}
\end{figure*}

\begin{figure}
\begin{tabular}{cc}
\subfloat[]{\includegraphics[scale=0.37]{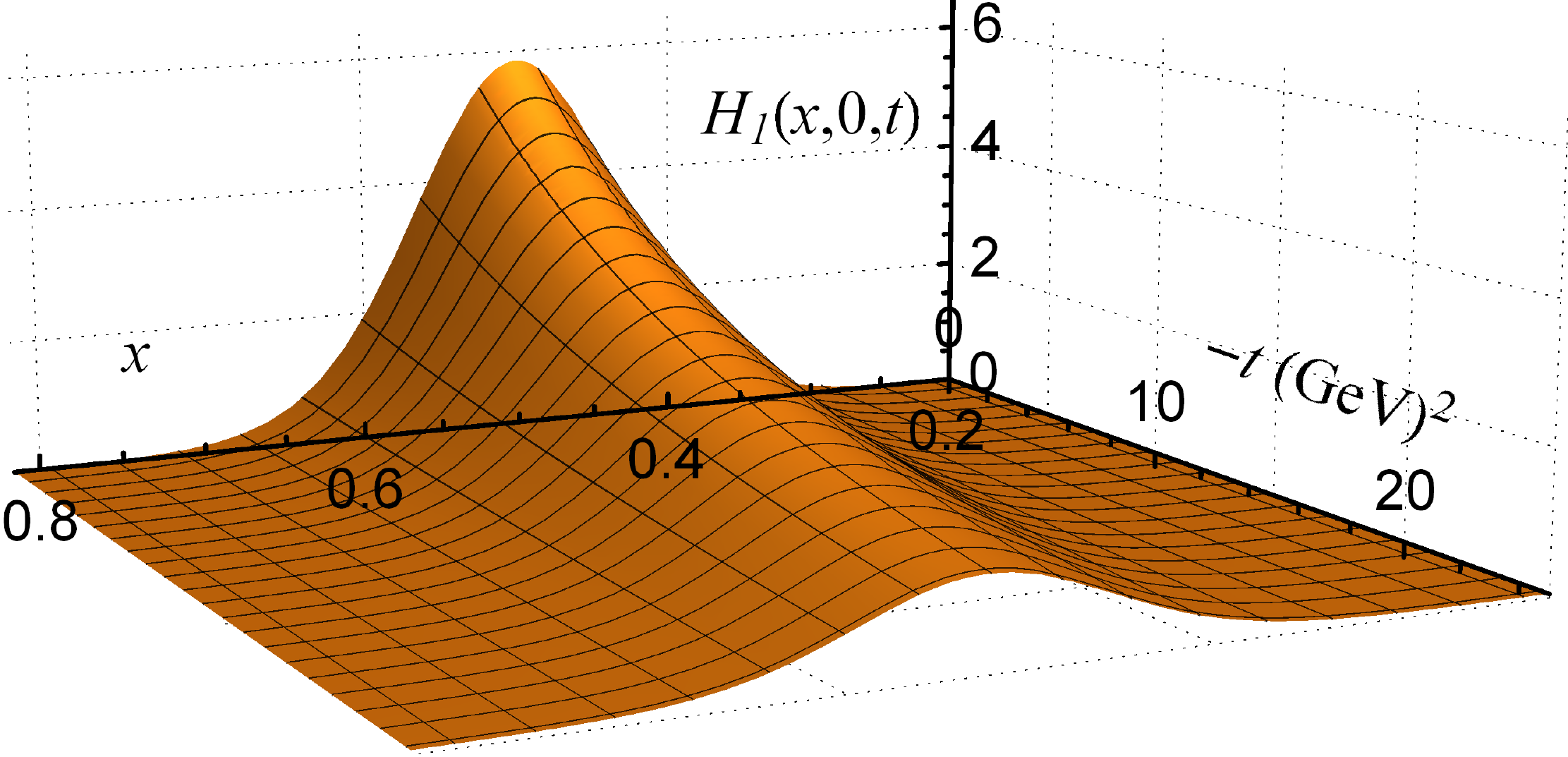}}\\
\subfloat[]{\includegraphics[scale=0.37]{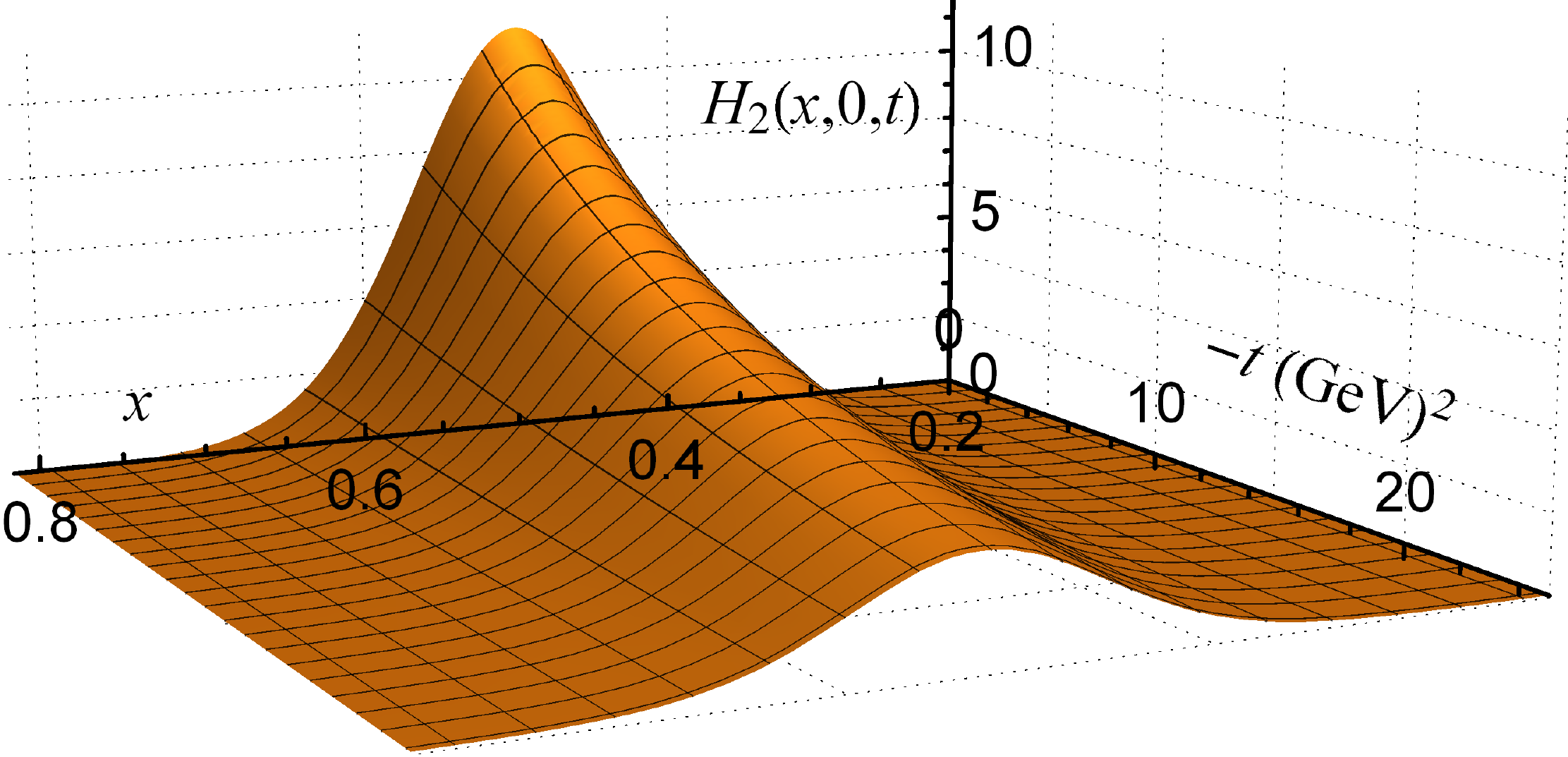}} \\
\end{tabular}
\begin{tabular}{cc}
\subfloat[]{\includegraphics[scale=0.37]{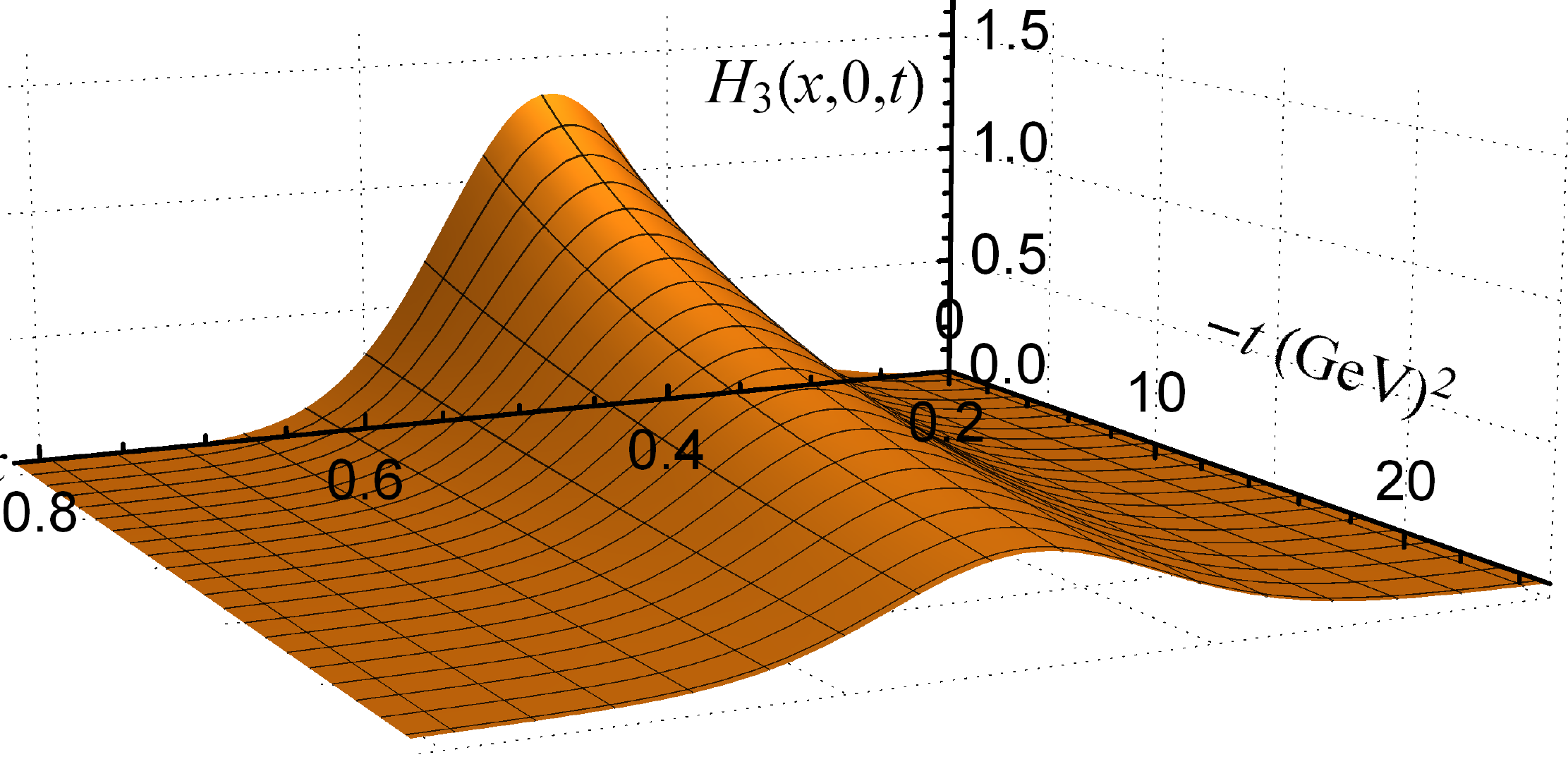}}\\
\subfloat[]{\includegraphics[scale=0.37]{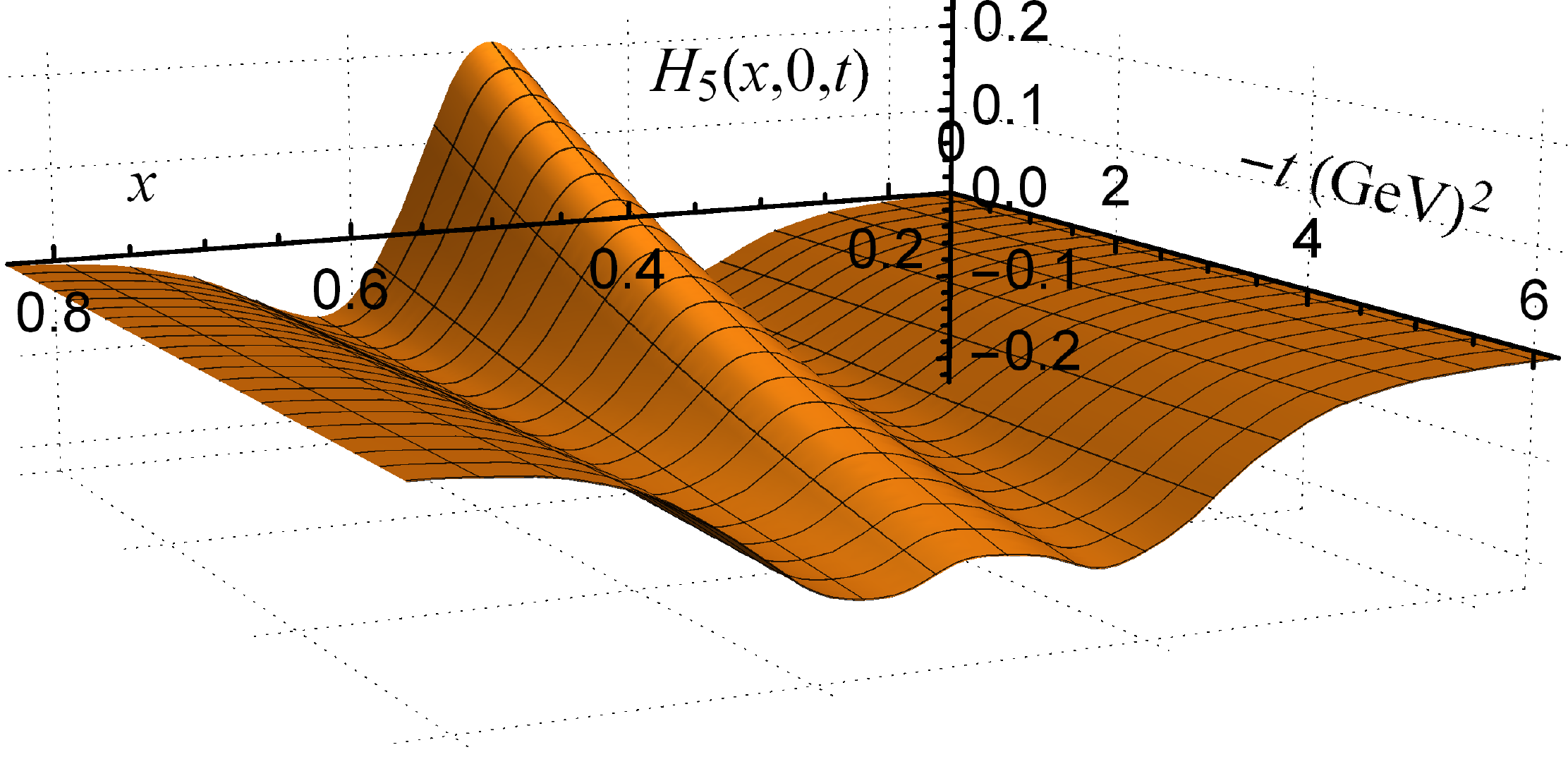}} \\
\end{tabular}
\caption{3D plot of helicity non-flip vector meson GPDs,  $H_i(x, \xi=0, t=-\Delta_\perp^2)$, $i=1, 2, 3, 5$ [Eqs.~\eqref{eq:gpd_amplitude1},~\eqref{eq:gpd_amplitude2},~\eqref{eq:gpd_amplitude3}, and~\eqref{eq:gpd_amplitude5}] for  $\Upsilon \, (1^3S_1)$ in the BLFQ approach. \label{fig:gpds_upsilon}}
\end{figure}

\begin{figure}
\begin{tabular}{cc}
\subfloat[]{\includegraphics[scale=0.37]{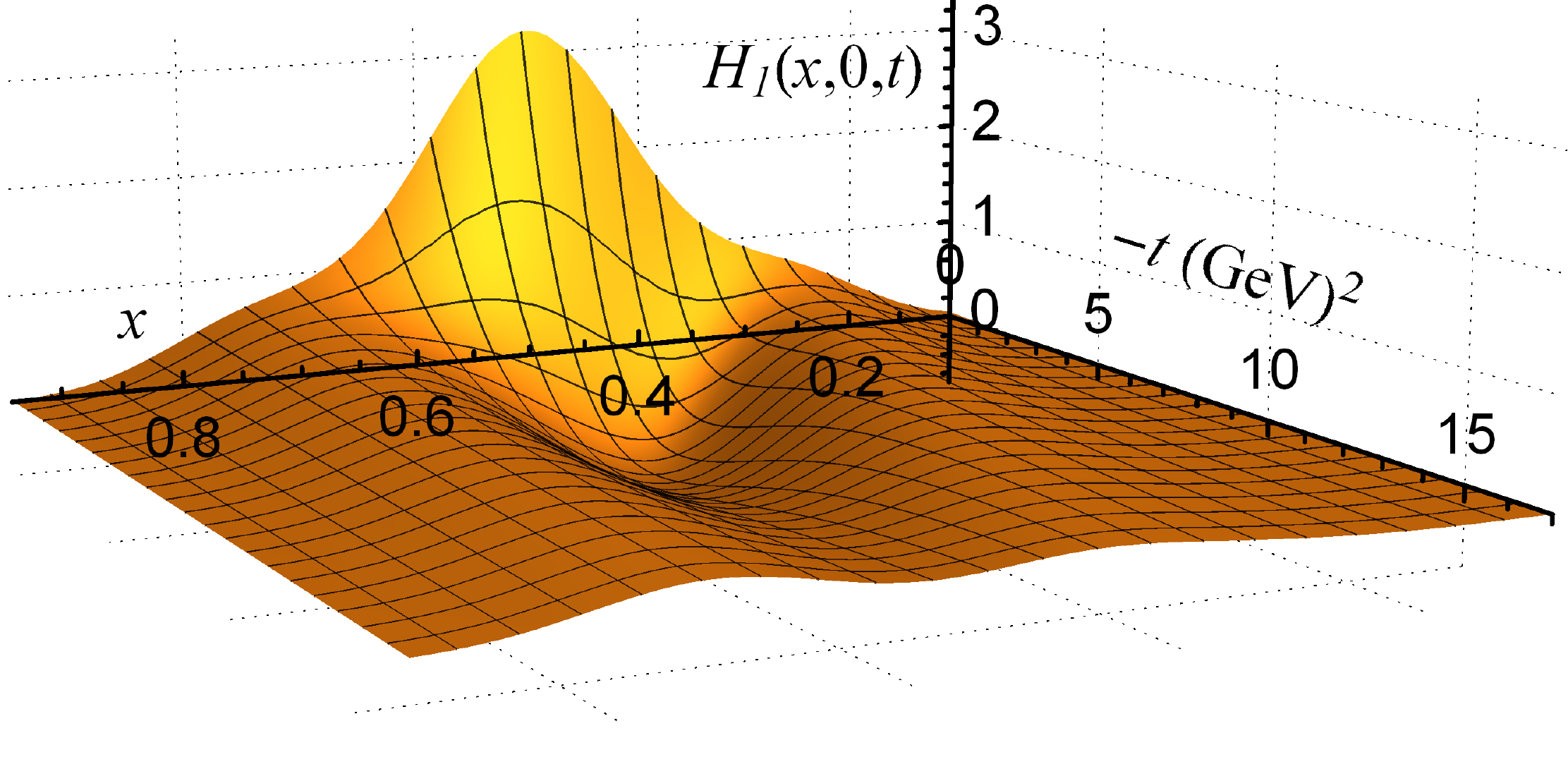}}\\
\subfloat[]{\includegraphics[scale=0.37]{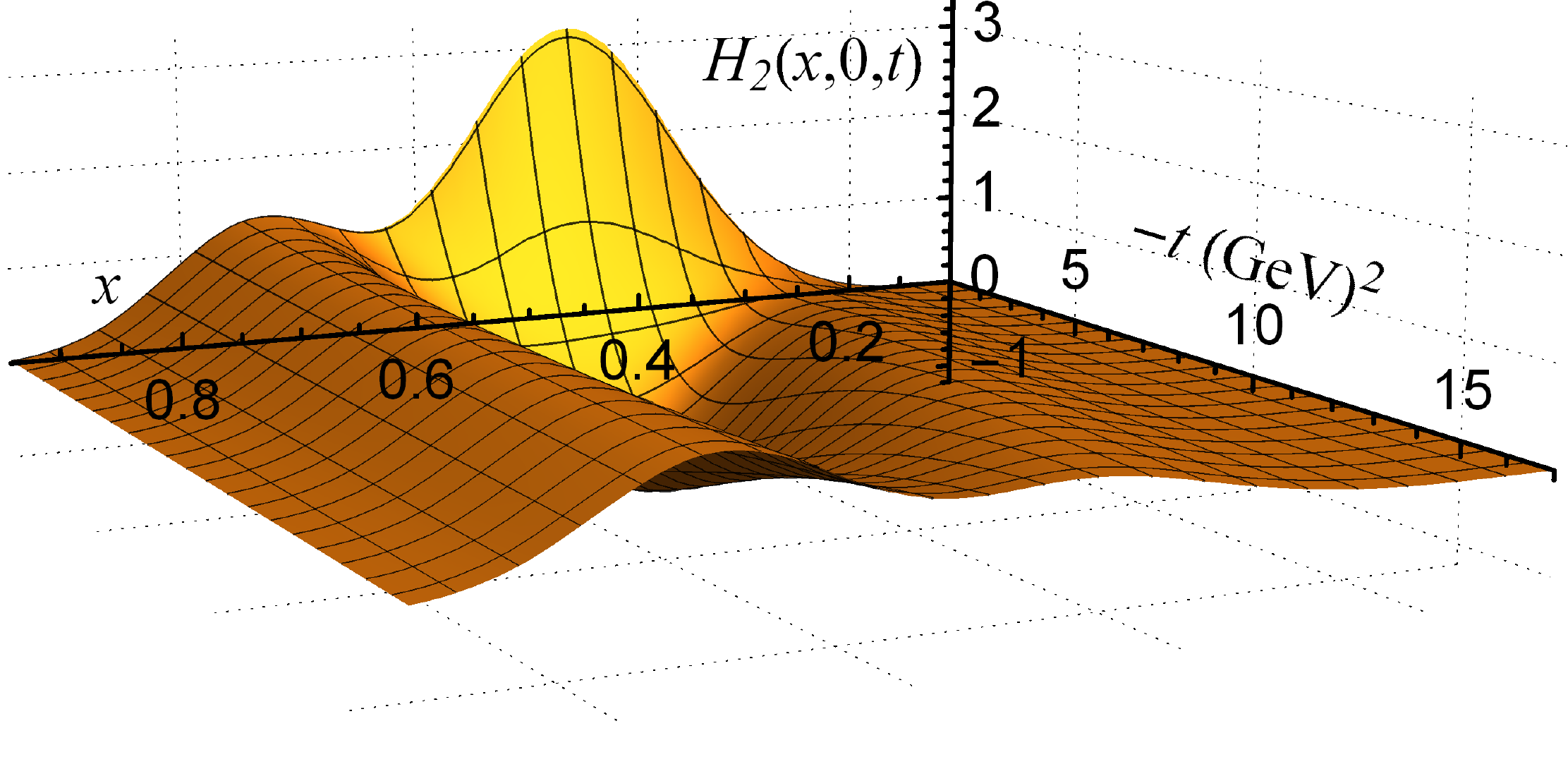}} \\
\end{tabular}
\begin{tabular}{cc}
\subfloat[]{\includegraphics[scale=0.37]{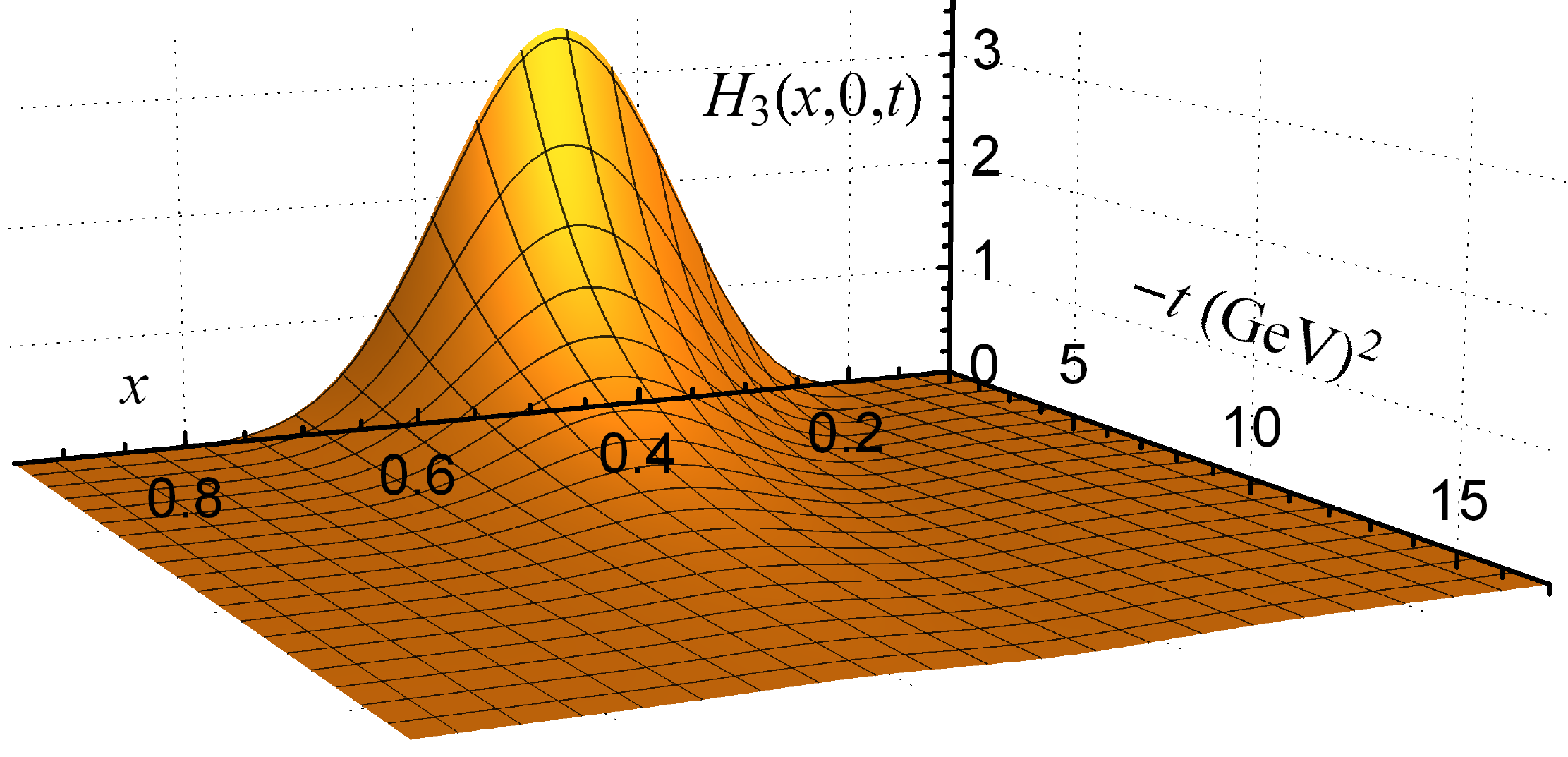}}\\
\subfloat[]{\includegraphics[scale=0.37]{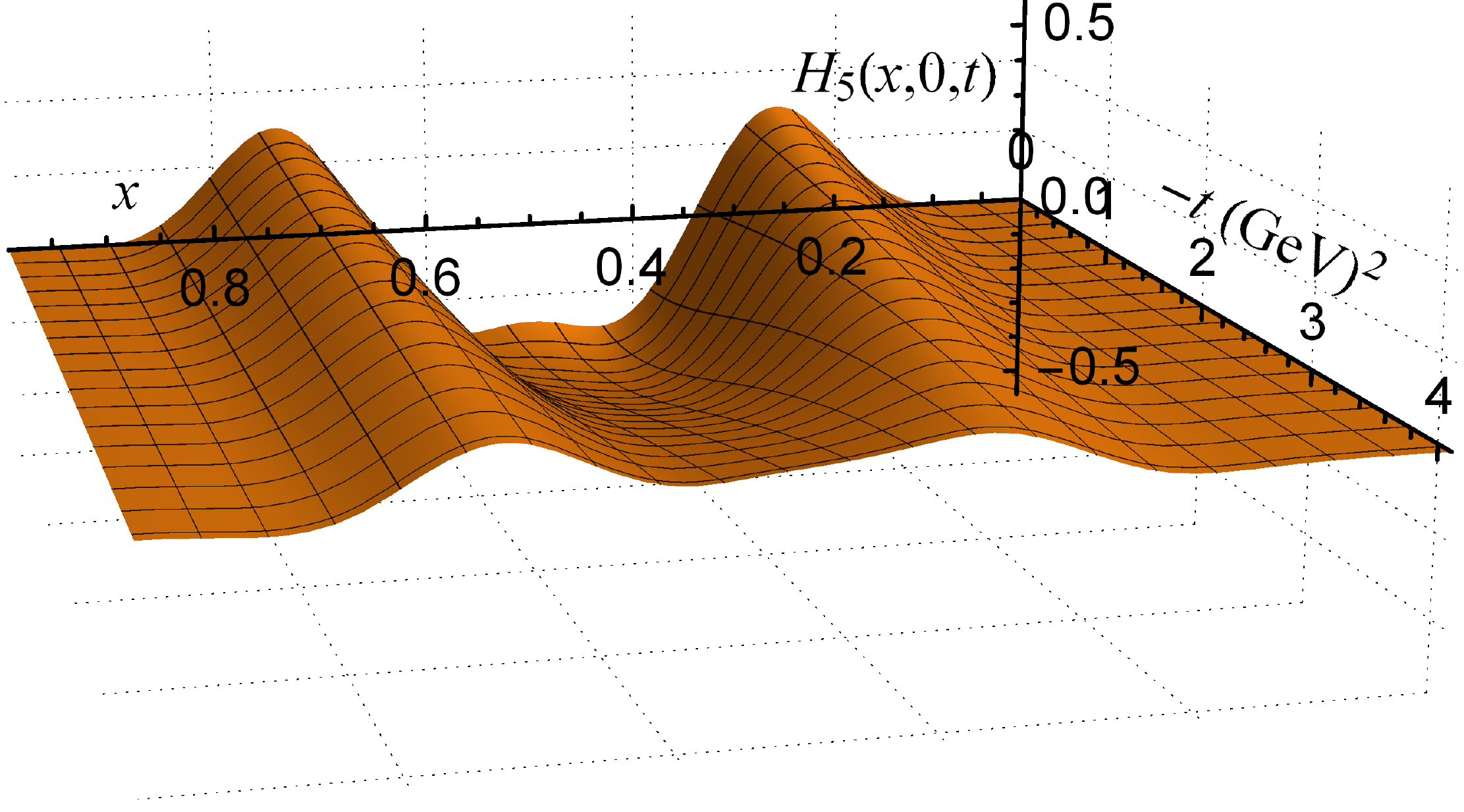}} \\
\end{tabular}
\caption{3D plot of helicity non-flip vector GPDs,  $H_i(x, \xi=0, t=-\Delta_\perp^2)$, $i=1, 2, 3, 5$ [Eqs.~\eqref{eq:gpd_amplitude1},~\eqref{eq:gpd_amplitude2},~\eqref{eq:gpd_amplitude3}, and~\eqref{eq:gpd_amplitude5}] for  $\psi^\prime \, (2^3S_1)$ in the BLFQ approach.\label{fig:gpds_psi}}
\end{figure}

\begin{figure}
\begin{tabular}{cc}
\subfloat[]{\includegraphics[scale=0.37]{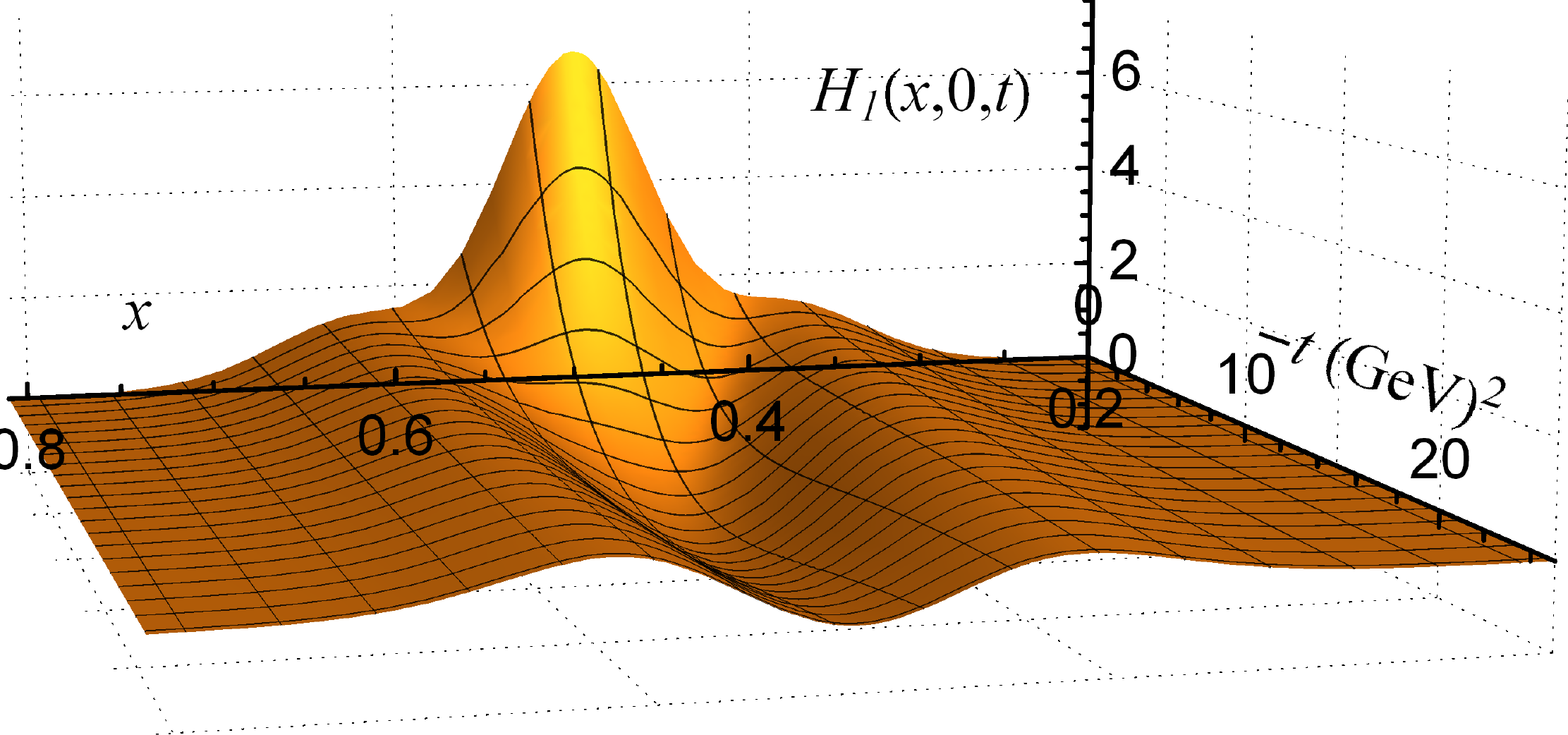}}\\
\subfloat[]{\includegraphics[scale=0.37]{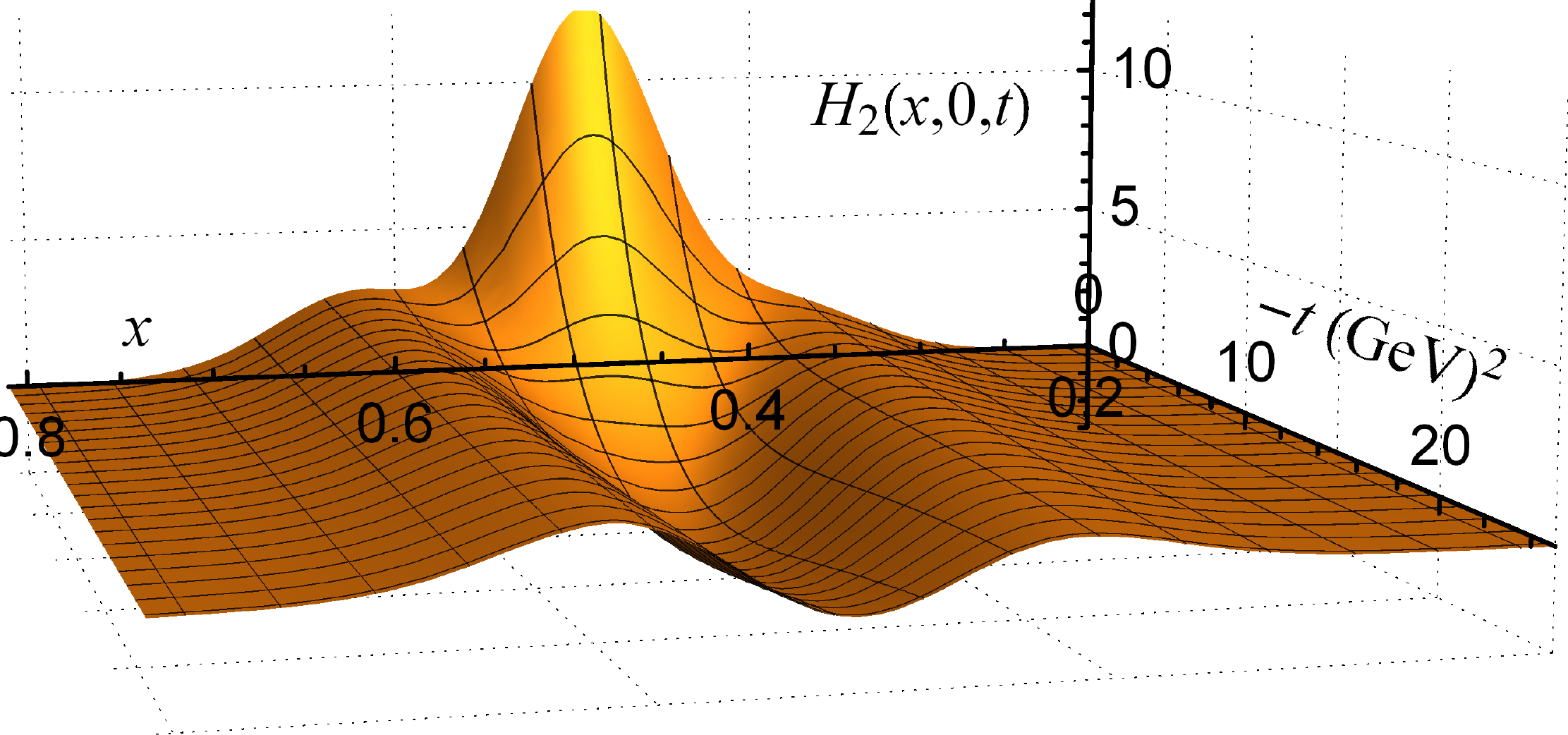}} \\
\end{tabular}
\begin{tabular}{cc}
\subfloat[]{\includegraphics[scale=0.37]{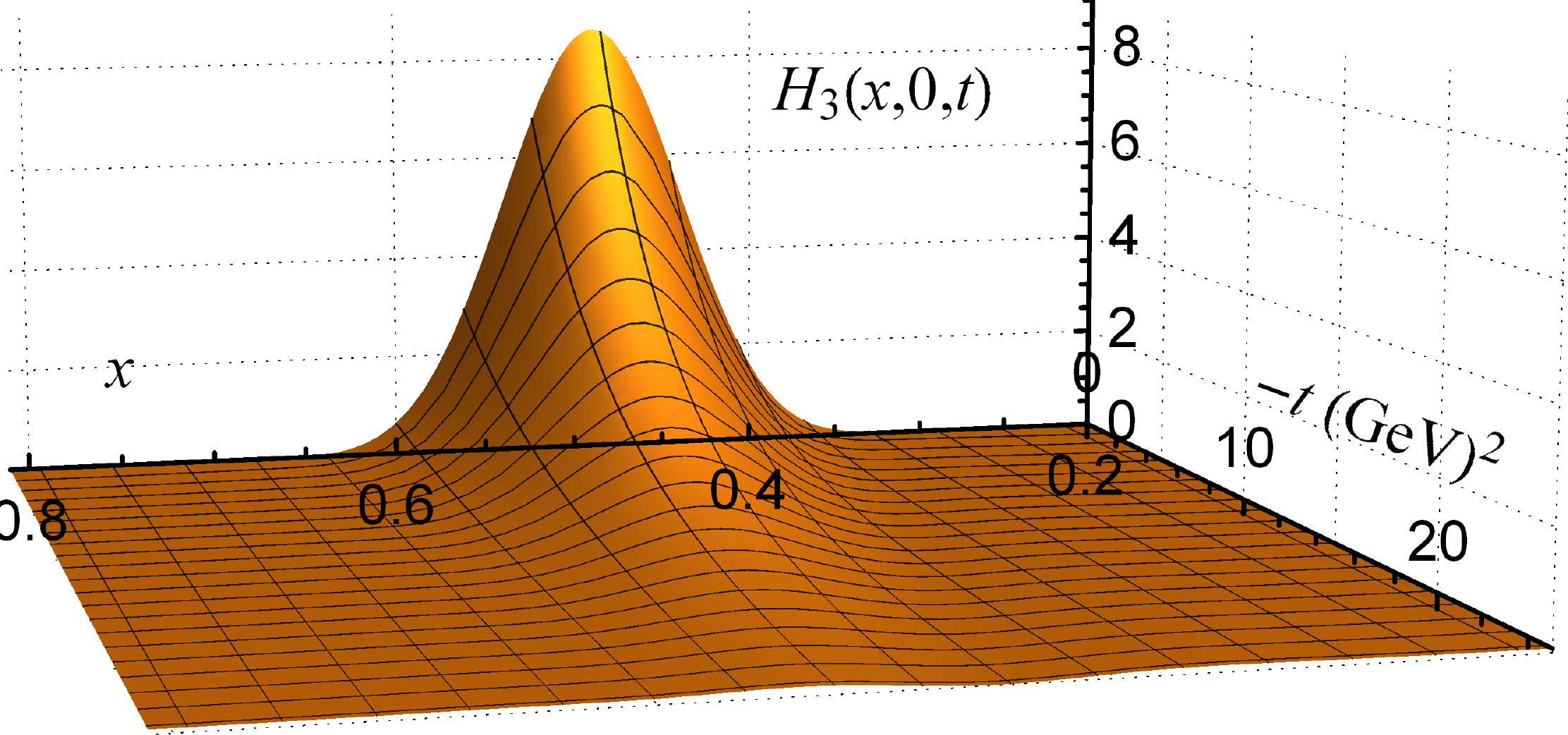}}\\
\subfloat[]{\includegraphics[scale=0.37]{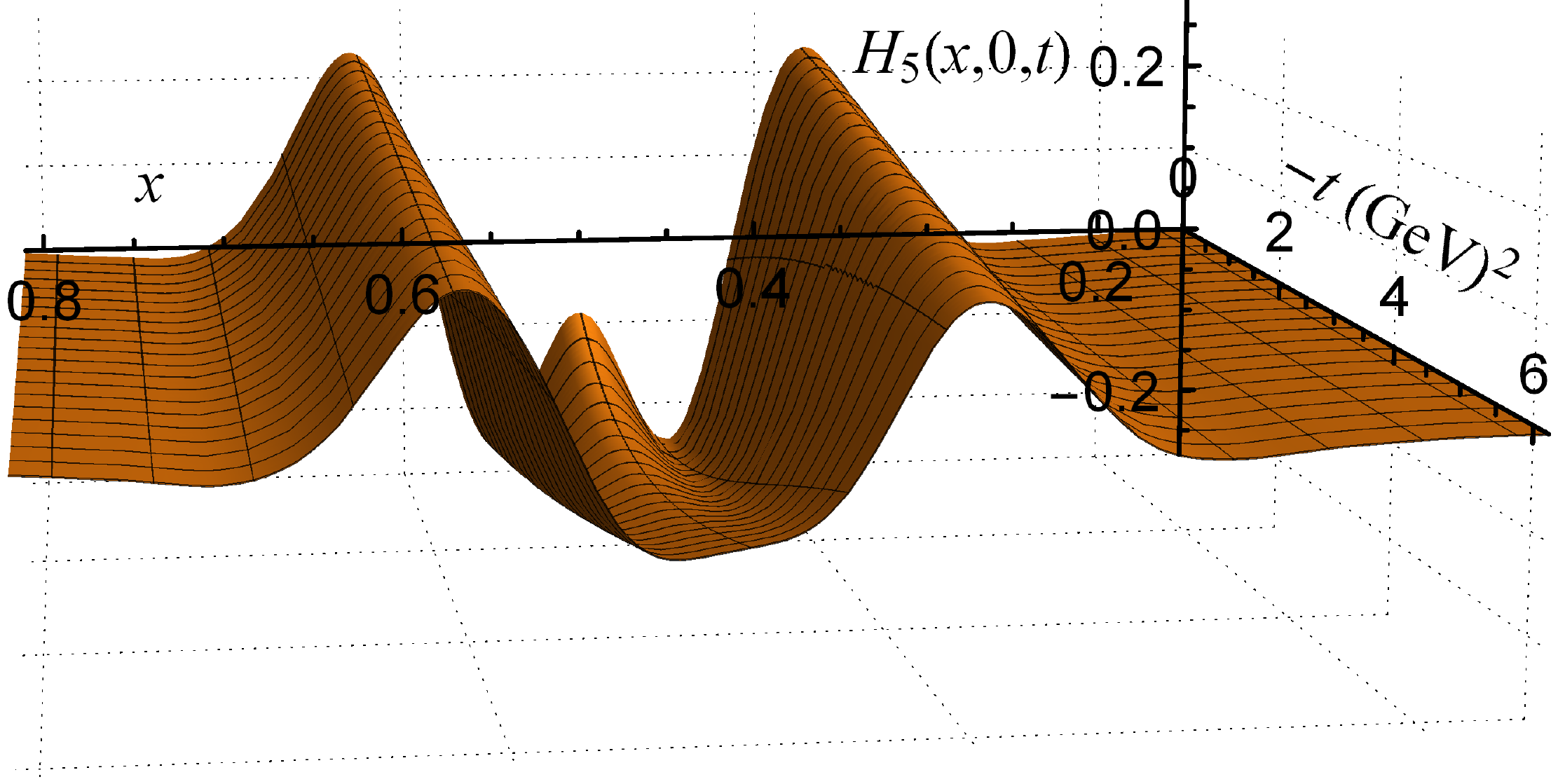}} \\
\end{tabular}
\caption{3D plot of helicity non-flip vector GPDs,  $H_i(x, \xi=0, t=-\Delta_\perp^2)$, $i=1, 2, 3, 5$ (Eqs.~\ref{eq:gpd_amplitude1},~\ref{eq:gpd_amplitude2},~\ref{eq:gpd_amplitude3},~\ref{eq:gpd_amplitude5}) for  $\Upsilon^\prime \, (2^3S_1)$  in the BLFQ approach. \label{fig:gpds_upsilon2S}}
\end{figure}
Let us turn our attention to vector meson GPDs for the ground state identified as $1^3S_1$ of heavy quarkonia\footnote{We use $N^{2S+1}S_J$ to identify meson states wherever relevant, where $N$ is the principal quantum number. According to the conventions of the Particle Data Group~\cite{Agashe:2014kda}, $N=L+n$, $n$ is the radial quantum number, $L$ is the total orbital quantum number, $S$ is the total intrinsic spin, and $J$ is the total angular momentum.} in the BLFQ approach with $N_{\text{max}}=L_{\text{max}}=24$. Each vector meson has five GPDs, but  we have seen from Eq.~\eqref{eq:gpd_amplitude4} that only four of them are non-zero on the LF. Figures~\ref{fig:gpds_jpsi} and ~\ref{fig:gpds_upsilon} show the non-zero GPDs for $J/\psi$ and $\Upsilon$, respectively. It is interesting to note that for both $J/\psi$ and $\Upsilon$, the integration of GPD $H_5(x, 0, t)$ over $x$ does not vanish except at $|t|=0$ which contradicts the consequence of Lorentz-invariance [see Eq.~\eqref{eq:H45zero} and associated text]. As we have pointed out before, the $x$-integration of GPD $H_5(x, 0, t)$ is the angular condition, the same term that is widely used in spin-one FF on the LF. It is not surprising that due to Lorentz symmetry breaking, the angular condition is not satisfied since we obtain a non-vanishing result for the $x$-integration of GPD $H_5(x, 0, t)$ except at the forward limit ($ |t| =0$). Repairing this deficiency requires the inclusion of higher Fock sectors in BLFQ, which is a subject for future research. The $x$-dependence of the ground-state vector meson GPDs is comparable with the corresponding quantities presented in Ref.~\cite{Sun:2017gtz}. The work in Ref.~\cite{Sun:2017gtz} presents the GPDs for the charged $\rho$ meson in a light front constituent quark model. Although the $\rho$ meson is light compared to the heavy quarkonia, the  peak somewhere between $x=0.4$ to $x= 0.6$ is the feature that both results have in common.

Next, we present GPDs for the radially excited  meson state identified as $2^3S_1$ ($\psi^\prime$ and $\Upsilon^\prime$). Figures~\ref{fig:gpds_psi} and ~\ref{fig:gpds_upsilon2S} show the non-zero vector meson GPDs for $\psi^\prime$ and $\Upsilon^\prime$, respectively. We again comment that, for both $\psi^\prime$ and $\Upsilon^\prime$, the integration of GPD $H_5(x, 0, t)$ over $x$ does not vanish except at $|t|=0$ similar to what we found for the case of ground-state vector meson. We note that the decaying trend of the vector meson GPDs  (Figs.~\ref{fig:gpds_psi},~\ref{fig:gpds_upsilon2S}) is more rapid with increasing $|t|$ for the radially excited state compared to the corresponding GPDs for the ground state (Figs.~\ref{fig:gpds_jpsi},~\ref{fig:gpds_upsilon}). This trend is consistent with the fact that $\psi^\prime$ ($\Upsilon^\prime$) has narrower radial extension in momentum space compared to that of $J/\psi$ ($\Upsilon$) . Similarly, these differences correlate with the relative sizes of these mesons as seen in the results of Table~\ref{tab:radii_vector}.  That is, larger charge r.m.s. radii correlate with smaller spread in momentum space, as expected. Furthermore, in the forward limit $t=0$,  the $x$-dependence of the vector meson GPDs  changes character significantly for  the radially excited states compared to that of the corresponding ground states. This observation is useful as the $x$-dependence of the GPDs in the forward limit is directly connected to the partonic interpretation of the hadronic spin~\cite{Hoodbhoy:1988am,Berger:2001zb,Taneja:2011sy}.

From the GPDs presented in this work it can be observed the decaying trend of the vector meson GPDs with $x$ is rapid for bottomonia compared to the GPDs for their counterparts in charmonia, and this trend can be understood  from considering the relative proximity to the non-relativistic limit where we expect that increasing quark mass leads to a sharper peak in $x$. Similarly, the rapid fall-off trend of the  GPDs  with $x$ for heavy quarkonia reflects the consequence of the impulse approximation. This follows the notion that the single quark cannot account for very large longitudinal momentum fraction for equal mass quark constituents. This observation is consistent with properties of the  deuteron vector GPDs available in Ref.~\cite{Cano:2003ju}.

Note the vector meson GPDs, investigated in this work, play important roles in various applications. The second moment of GPD $H_2(x,0,0)$ gives the spin-one angular momentum via a sum rule~\cite{Taneja:2011sy,Abidin:2008ku}, and GPD $H_5(x,0,0)$ is equal to $b_1(x)$, the Deep Inelastic Scattering (DIS) structure function, for the spin-one target such as the deuteron~\cite{Liuti:2014dda,Taneja:2011sy,Berger:2001zb,Close:1990zw}. There is growing interest in these quantities since the announcements of the experimental measurements on $b_1(x)$  from HERMES ~\cite{Airapetian:2001yk,Hoodbhoy:1988am}. Our GPD results, presented in the off-forward limit in this work provide insight to further investigate angular momentum and the structure functions for the spin-one target within BLFQ.

\section{summary and outlook \label{Summary}}
We have calculated the EM FFs for a selection of heavy quarkonia. We have compared the charge radii, the magnetic moments and quadrupole moments calculated  in both BLFQ and SBL approaches with the results from other approaches available in the literature. The differences between the BLFQ and SBL results for selected mesons highlight the dynamics of the internal structure of heavy quarkonia. We have also studied the convergence of BLFQ results and find  good convergence at $N_{\text{max}}=L_{\text{max}}=24$. We presented the GPDs for selected (pseudo) scalar and vector mesons.  We have also pointed out that our GPD results in specific kinematic regions, can be linked with DIS structure functions to further investigate the spin-one hadronic structure. Furthermore, our GPD results in three dimensions, in the region $t\neq 0$ provide insight into the non-perturbative structure of the spin-one system and  could facilitate making connections between GPDs and the  partonic interpretation in the off-forward limit.

We foresee a number of extensions such as the adoption of BLFQ results with running coupling~\cite{Li:2017mlw}. In addition, within BLFQ, one can choose the transverse component of the current operator to calculate the magnetic form factor and compare the associated results with the corresponding quantities presented in this work~\cite{Li:2018uif}. This proposal is inspired by the fact that, in non-relativistic quantum mechanics, the magnetic moments are computed from the spatial current density operator. One can further investigate GPDs in the non-zero longitudinal momentum transfer frame. Within BLFQ,  the gravitational form factors of the spin-one hadrons can be studied via the second moment of such GPDs. Such an investigation within BLFQ can provide insight to the hadronic spin structure, and in particular the quark's angular momentum within the hadron. In addition, within this formalism, one can calculate the transverse momentum dependent distribution (TMDs) for spin-one mesons and investigate the meson spin contribution that is carried by orbital angular momentum of the quarks. Ultimately, within the BLFQ approach, one can investigate proton's spin structure and reveal the dynamics of quark and gluon contributions to the proton spin.
\section{acknowledgments}
We gratefully acknowledge valuable discussions with P. Maris. This work was supported in part by the US Department of Energy (DOE) under Grants Nos. DE-FG0287ER40371 and DE-SC0018223 (SciDAC-4/NUCLEI). Computational resources were provided by the National Energy Research Scientific Computing Center (NERSC), which is supported by the Office of Science of the US DOE under Contract No. DE-AC02-05CH11231.

\end{document}